\RequirePackage{fix-cm}
\documentclass[smallextended]{svjour3}       % onecolumn (second format)
\smartqed  % flush right qed marks, e.g. at end of proof
\usepackage{graphicx}
\usepackage{cite}

\usepackage{ocgx2}

\begin{document}

\title{Spectral engineering via complex patterns of circular nano-object miniarrays: I. convex patterns tunable by integrated lithography realized by circularly polarized light%\thanks{Grants or other notes
}

\titlerunning{Spectral engineering via complex convex patterns}        % if too long for running head

\author{Aron Sipos          \and
        Emese Toth			\and
        Oliver A. Fekete 	\and
        Maria Csete* %etc.
}


\institute{Aron Sipos \at
			  Institute of Biophysics, Biological Research Centre, \\
			  H-6726 Szeged Temesvári krt. 62, Hungary \\
              \email{sipos.aron@brc.hu}           %  \\
           \and
           Emese Toth \at
              Department of Optics and Quantum Electronics, \\
              University of Szeged, H-6720, Szeged Dóm tér 9, Hungary \\
%              \email{fauthor@example.com}           %  \\
           \and
           Oliver A. Fekete \at
              Department of Optics and Quantum Electronics, \\
              University of Szeged, H-6720, Szeged Dóm tér 9, Hungary \\
%              \email{fauthor@example.com}           %  \\
           \and
           *Maria Csete \at
              Department of Optics and Quantum Electronics, \\
              University of Szeged, H-6720, Szeged Dóm tér 9, Hungary \\
              \email{mcsete@physx.u-szeged.hu}           %  \\              
}

%\date{ }

\maketitle

\begin{abstract}

Illumination of colloid sphere monolayers by circularly polarized beams enables the fabrication of concave patterns consisting of circular nanohole miniarrays that can be transferred into convex metal nanoparticle patterns via a lift-off procedure. Unique spectral and near-field properties are achievable by tuning the geometry of the central nanoring and quadrumer of slightly rotated satellite nanocrescents and by selecting those azimuthal orientations that promote localized plasmon resonances. The spectral and near-field effects of hexagonal patterns made of uniform gold nanorings and nanocrescents, that can be prepared by transferring masks fabricated by a perpendicularly and obliquely incident single homogeneous circularly polarized beam, were studied to uncover the supported localized plasmonic modes. Artificial rectangular patterns made of a singlet nanoring and singlet nanocrescent as well as quadrumer of four nanocrescents were investigated to analyze the role of nano-object interactions and lattice type. It was proven that all nanophotonical phenomena are governed by the azimuthal orientation independent localized resonance on the nanorings and by the C2, C1 and U resonances on the nanocrescents in case of $\bar{E}$-field oscillation direction perpendicular and parallel to their symmetry axes. The interaction between localized surface plasmon resonances on individual nano-objects is weak, whereas scattered photonic modes have a perturbative role at the Rayleigh anomaly only on the larger periodic rectangular pattern of miniarrays. Considerable fluorescence enhancement of dipolar emitters is achievable at spectral locations promoting the C and U resonances on the composing nano-objects.
\end{abstract}

\section{Introduction}
\label{intro}

Investigation of individual and periodic plasmonic structures initialized the inspection of complex patterns made of metal nano-objects, which have interesting synergetic near-field and spectral properties. The building blocks of these structures are usually uniform nano-objects, and the array optical properties can be tuned by varying the geometrical parameters of the composing nano-objects and the periodicity of their pattern \cite{ref_01,ref_02,ref_03,ref_04}.\\
The optical properties of individual convex scatterers depend on their size, shape and are influenced by the dielectric environment as well \cite{ref_05, ref_06}. The extinction spectrum of small spherical particles is dominated by a dipolar resonance, while on larger particles multipolar modes develop as well \cite{ref_07}. Plasmon hybridization on spherical nano-particles with finite metal walls, like rings and shells, makes it possible to attain larger number of multipolar modes that are tunable by varying the geometrical parameters \cite{ref_08}. The symmetric / antisymmetric modes result in peaks, which are red / blue-shifted by decreasing the wall thickness to core ratio \cite{ref_09,ref_10}. \\
On dimers (trimers) of core-shells electric (magnetic) dipoles develop, while on heptamer miniarrays Fano-like peaks appear due to the coexistent bright and dark modes originating from parallel and anti-parallel dipoles \cite{ref_11}. Asymmetric plasmonic trimer structures exhibit a broadband polarization independent far-field scattering and enhanced near-field simultaneously \cite{ref_12}.\\
The Fano-peaks can be precisely tuned on quadrumers and in asymmetric heptamers by the azimuthal orientation, as well as on hetero-octamers by the geometrical parameters \cite{ref_13,ref_14}. Oligomers built up from a nanocross and a nanorod act as an ultracompact directional antenna at the Fano resonance, whereas hetero-heptamers built up from a central nanocross and six satellite rod-pairs result in multiple Fano-lines \cite{ref_15}.\\
It was shown that hexagonal and square patterns made of hemi-ellipsoids are capable of resulting in band-gaps, which can be optimized by the geometrical parameters \cite{ref_16}. On hexagonal and rectangular arrays of cylinders and prisms dipolar coupling results in a distance dependent spectral shift \cite{ref_17}. The dispersion characteristics of hexagonal patterns made of convex nanoparticles is governed mainly by LSPR, in contrast to nanoholes in metal films, where the diffractive coupling on the pattern results in characteristic propagating plasmon related bands \cite{ref_18}. The diffractive coupling can result in sharp peaks on the extinction of convex patterns, when the lattice period is tuned to the LSPR of individual nano-particles, which phenomenon is the surface lattice resonance (SLR) \cite{ref_01,ref_02,ref_03,ref_19,ref_20}. The interaction of localized plasmon resonance and diffraction orders results in narrow peaks on the normalized near-field as well \cite{ref_21}. Moreover, adding randomness to rectangular arrays of nano-particles makes it possible to realize fine spectral engineering and to achieve further enhanced $\bar{E}$-field \cite{ref_22}. \\
On randomly arranged C-shaped gold nanocrescents prepared by shadow sphere lithography the polarization, opening angle, thickness and diameter dependence of the resonances was demonstrated \cite{ref_23}. The C and U resonances excited by light polarized perpendicular and parallel to the symmetry axes were identified as antenna-like resonances. The polarization dependency of multipoles, accompanying $\bar{E}$-field confinement and corresponding resonance spectra was demonstrated on individual C-shaped SSRs as well \cite{ref_24}. \\
On hexagonal arrays of C-shaped resonators created by shadow sphere lithography the strong effect of the opening angle as well as of the polarization on the parity and strength of excitable plasmonic modes was demonstrated \cite{ref_25}. Caused by the finite scaling regime of magnetic resonances on gold, application of Al was suggested \cite{ref_26,ref_27}.\\ 
Complex plasmonic materials built up from varieties of split ring resonators (SRR) can act as metamaterials as well. Single ring multi-cut building blocks in rectangular arrays with unit cells in the order of 100 nm are capable of shifting the operation regions into the visible regime \cite{ref_28}. Cavity resonances excitable on SRR arrays can possess high quality factors \cite{ref_29}. In rectangular-patterns of U-shaped SRR doublets the asymmetry of coupled elements makes it possible to excite dark modes resulting in narrow lines, which is advantageous in sensing \cite{ref_30}. Mirror symmetry broken metamaterials show Fano resonances of controllable width that can be tuned by the temperature \cite{ref_31}.\\
Arrays of various plasmonic objects have been extensively applied in biosensing. Hexagonal patterns consisting of nano-tetrahedra and spheroids prepared by nano-sphere lithography (NSL) have shown bio-sensing capabilities \cite{ref_06}. Short range ordered disks and oblate spheroids prepared by NSL exhibit a sensitivity dependent on the individual object geometry \cite{ref_32}. Hexagonal arrays of nanocrescents are capable of promoting SERS \cite{ref_33}. Via plasmonic nano-disk-arrays six orders of magnitude fluorescence enhancement was observed on pre-designed chips \cite{ref_34}. Two orders of magnitude spontaneous emission enhancement has been achieved in case of dipolar emitters oriented perpendicular to the torus axes that are arranged on gratings \cite{ref_35}. This example indicates that complex patterns composed of predesigned rings can result in considerable enhancement. The highest fluorescence efficiency can be achieved by positioning tapered antennas like spheres, disks and tori above thin metals films \cite{ref_36}. These results predict that combination of thin films and spherical antennas is advantageous in sensing of dipolar emitters.\\
To fabricate complex patterns various methodologies have been already developed. Via shadow sphere lithography (SSL) randomly distributed nanocrescents can be fabricated by evaporating metal layers onto surfaces consisting of randomly distributed colloids \cite{ref_23}. Colloid sphere lithography (CSL) makes it possible to generate various nano-objects (cups, rods and wires), however all ordered patterns inherit the hexagonal symmetry of the closely packed monolayers \cite{ref_37,ref_38}. Complex structures with hexagonal symmetry including various building blocks and multiple materials can be fabricated via sequential deposition from multiple angles in CSL \cite{ref_39}. Individual colloid spheres or monolayers of them can be used as masks as well. Special methodology has been developed to fabricate randomly distributed nanorings on surfaces, which exhibit spectral peaks in regimes not accessible via disks, moreover support special magnetic states, when made of ferromagnetic materials \cite{ref_09,ref_40}. Edge spreading lithography is capable of creating double rings made of gold with a thickness on the order of 10 nm \cite{ref_41,ref_42}.\\
Special multistep methodologies have been developed to break the hexagonal symmetry and to overcome the monodispersity achievable via colloid sphere monolayers. To fabricate hierarchical structures, CSL was combined with RIE and annealing as well as tilted evaporation, photo-masks projection, metal sputtering and ion milling, the latter is capable of resulting in nanorings \cite{ref_43}. However, in case of photomask projections, the periodicity of the mask was significantly larger than the colloid-sphere diameter in the illuminated monolayer. \\
Deposition of gold nanoparticles was performed onto periodic patterns created by two-beam interference on colloidal crystals, where photo-switching in the medium promotes chemical attachment \cite{ref_44}. The period of the interference pattern was one and two orders of magnitude larger, than the diameter of spheres forming the monolayer. Binary colloid monolayers were created with characteristic size-scale ratio on the order of 0.19-0.4, however the hexagonal symmetry was inherited also in this case \cite{ref_45}. \\
In this paper we present complex structures, which are composed of convex miniarrays of a central nanoring and satellite nanocrescents, arrayed in a rectangular pattern. These structures can be fabricated by combining colloid sphere and interference lithography, in a two-steps procedure \cite{ref_46,ref_47,ref_48,ref_49}. The interferometric illumination of colloid-sphere monolayers (IICSM) enables the direct structuring of positive resists, whereas the masks can be transferred into convex patterns by a lift-off procedure. The spectral and near-field properties of complex concave patterns consisting of rounded nanohole miniarrays is described in Part II of this paper, where it is detailed, that the rounded nano-objects originate from circularly polarized light, whereas the pattern periodicity is determined by the interference pattern periodicity \cite{ref_50}. In order to uncover the localized modes supported by the building blocks, the spectral and near-field effects of hexagonal patterns composed of uniform nanorings and nanocrescents that can be fabricated by colloid sphere lithography followed by a lift-off procedure, was also studied. To analyze the role of nano-object interactions and the lattice type in the collective resonance, artificial rectangular patterns of a singlet nanoring and singlet nanocrescent as well as quadrumer of four nanocrescents were investigated. A comparative study about the spectral and near-field effects of concave and convex patterns, that are achievable via IICSM directly and via a consecutive lift-off procedure, is provided as well in our previous work and in an upcoming paper \cite{ref_49,ref_51}.

\section{Method}
\label{Met}

%Text with citations \cite{ref_01} and \cite{ref_02}.

\subsection{Numerical modeling and characterization of patterns consisting of convex spherical nano-objects}
\label{Num_mod}

\begin{figure}[h]
\center
	{\includegraphics[width=0.75\textwidth]{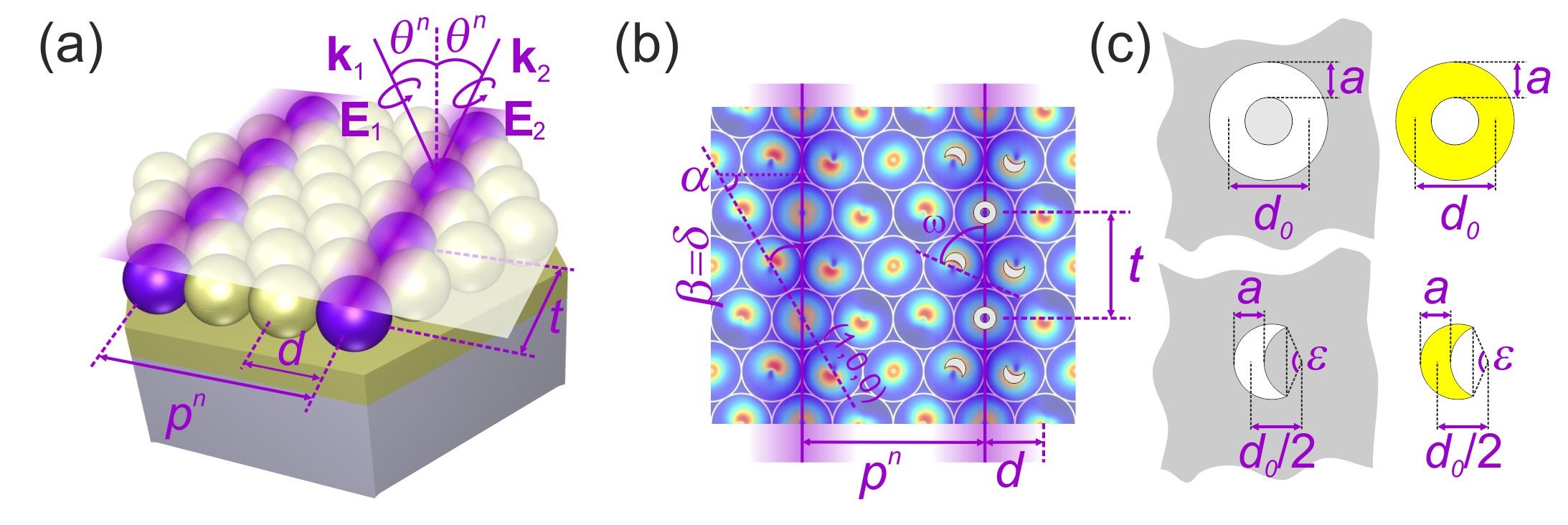}}
\caption{Schematic drawing of a multilayer that can be structured in IICSM methodology followed by a lift-off procedure. Characteristic size parameters of the resist pattern and of the transferred metal pattern (a) 3D scheme of $p^n$ pattern period variation by $\theta^n$ polar angle tuning (b) orientation of incidence plane ($\alpha$) and interference pattern ($\beta$) to tune colloid sphere $d$ diameter-scaled $t$ distance of nano-objects, (c) generated features qualified by size parameters of $d_0$, $a$, $\varepsilon$ gap angle and $\omega$ orientation.}
\label{colloid_fig_01}       % Give a unique label
\end{figure}

Finite element method (FEM) implemented into the Radio Frequency module of COMSOL Multiphysics software package (COMSOL AB, Sweden) was used to determine the geometrical parameters of convex patterns achievable with IICSM. The $\bar{E}$-field distribution was examined under the Au colloid sphere monolayer on the surface of the multilayers. In these studies, all materials were modeled by taking into account the wavelength dependent optical properties \cite{ref_52,ref_53}.\\
The schematic drawing in Fig. \ref{colloid_fig_01}a shows the multilayer applied for illumination by circularly polarized light, which treatment is followed by a lift-off procedure to convert the concave resist hole-pattern into an analogue convex metal patter. Fig. \ref{colloid_fig_01}b and c show the main characteristic geometrical parameters variable via IICSM methodology, when it is realized by circularly polarized beams. In the closely packed in between arrays illumination configuration the $p$ pattern periodicity is determined by the $\theta$ angle of incidence and $\lambda$ wavelength, and it can be tuned in half colloid sphere diameter $(d/2)$ steps. The $t$ inter-object distance is determined be the $\alpha$ and $\beta$ orientation of the incidence plane and of the interference pattern. The $d_0$ and $a$ nano-object size parameters are dependent not only on the wavelength and power density, colloid sphere $d$ diameter and material, but can be influenced by the lift-off-procedure as well.\\

\subsection{Spectral and near-field study of different patterns}
\label{Spe_nea_fie_stu}
FEM (COMSOL) was used to demonstrate the spectral engineering capabilities of convex patterns that can be fabricated by applying circularly polarized light in IICSM. Complete spectral study of these complex patterns was performed, namely the optical responses of patterns consisting of convex gold nano-objects with 45 nm height were determined (Fig. \ref{colloid_fig_02}). The inspected realistic hexagonal patterns are as follows: hexagonal pattern of (i) nanorings (Fig. \ref{IICSM_DG_cx_01}), (ii) horizontal nanocrescents (Fig. \ref{IICSM_DG_cx_01}, vertical nanocrescents are presented in a Supplementary material). The studied artificial composing rectangular patterns are as follows: 300 nm rectangular pattern of (iii) a singlet nanoring (Fig. \ref{IICSM_DG_cx_02}), (iv) a singlet horizontal nanocrescent (Fig. \ref{IICSM_DG_cx_03}), (v) a quadrumer of slightly rotated nanocrescents (Fig. \ref{IICSM_DG_cx_04}). Finally, two different complex rectangular patterns were analyzed: (vi) 300 nm (Fig. \ref{IICSM_DG_cx_05}) and (vii) 600 nm (Fig. \ref{IICSM_DG_cx_06}) rectangular pattern of the same miniarray composed of a central nanoring and quadrumer of slightly rotated nanocrescents. \\
To analyze the spectral and near-field properties, the convex patterns were re-illuminated by p-polarized light by varying $\gamma$ azimuthal angle, which qualifies the incidence plane orientation with respect to the y axes. In this spectral study p-polarized light illuminated the convex gold nano-objects in a symmetric environment, supposing that the nanoring and nanocrescents are surrounded by, and the nanorings' inner holes are filled with NBK7 medium. The hexagonal pattern of convex nanorings and nanocrescents have been inspected in $0^{\circ}$ and $90^{\circ}$ azimuthal orientations in order to uncover the characteristic LSPRs supported by the nano-objects. These $\gamma$ azimuthal angles ensure $\bar{E}$-field oscillation direction along ($0^{\circ}$($90^{\circ}$)) and perpendicularly ($90^{\circ}$($0^{\circ}$)) to the horizontal (vertical) nanocrescent symmetry axes, that orientations promote U and C type LSPR on crescent shaped nano-objects in accordance with the literature, respectively \cite{ref_23}. Both the $0^{\circ}$/$16^{\circ}$ and $90^{\circ}$/$106^{\circ}$ azimuthal orientations of rectangular patterns have been inspected, since these promote LSPR as well as surface lattice resonance effects in case of singlet horizontal nanocrescents / LSPR in cases of the quadrumer and either miniarrays. The spectra were taken throughout the 200 nm–1000 nm interval with 10 nm resolution, at $\varphi$ = $0^{\circ}$ polar angle corresponding to perpendicular incidence (Fig. \ref{IICSM_DG_cx_01}-\ref{IICSM_DG_cx_06}/a).\\
COMSOL was applied to inspect the dispersion characteristics of all convex patterns by selecting fractions on the high symmetry path throughout their IBZ according to the azimuthal orientations that promote uncovering of the LSPR on composing circular nano-objects and mapping of the photonic modes scattering. Accordingly, the dispersion characteristics have been taken in $0^{\circ}$ and $90^{\circ}$ azimuthal orientations. In case of dispersion diagram computations, the spectral range was extended through 1000 nm, applying the same 10 nm wavelength resolution as in case of perpendicular incidence, whereas the $\varphi$ incidence angle was also modified from $0^{\circ}$ to $85^{\circ}$ with $5^{\circ}$ steps (Fig. \ref{IICSM_DG_cx_01}d, e and Fig. \ref{IICSM_DG_cx_02}-\ref{IICSM_DG_cx_06}/c). Wherever needed to uncover all underlying modes on the dispersion graphs, higher resolution complementary calculations were performed with smaller steps.\\
According to the literature, in case of plasmonic patterns the absorptance spectra are the most informative to find resonances, therefore the absorptance spectra and the dispersion characteristics taken in p-polarized absorptance are analyzed throughout this paper \cite{ref_15,ref_54}. \\
During inspection of the near-field phenomena, first the characteristic charge distribution corresponding to C and U resonance has been determined based on previous results in the literature, then the accompanying EM-field has been inspected on the time-averaged $E_z$ distribution \cite{ref_23}. The latter promoted to determine the complementary EM-field and charge distributions on concave patterns (Fig.  \ref{IICSM_DG_cx_01}b, c and Fig. \ref{IICSM_DG_cx_02}-\ref{IICSM_DG_cx_06}/b) \cite{ref_49,ref_50,ref_51}.\\
Localised plasmon resonances on the nanorings and nanocrecents are distinguished by using “r” and “c” in the abbreviations.\\
FEM was used to inspect the capabilities of the artificial composing patterns, namely the rectangular patterns of singlet nanoring and quadrumer of nanocrescents, as well as of the rectangular pattern of their miniarray to enhance the fluorescence (Fig. \ref{dip_encha_01}).

\section{Results and discussion}
\label{Res_dis}

\subsection{Patterns achievable in different illumination configurations}
\label{Pat_in_dif_ill}

\begin{figure}[h]
\center
	{\includegraphics[width=0.75\textwidth]{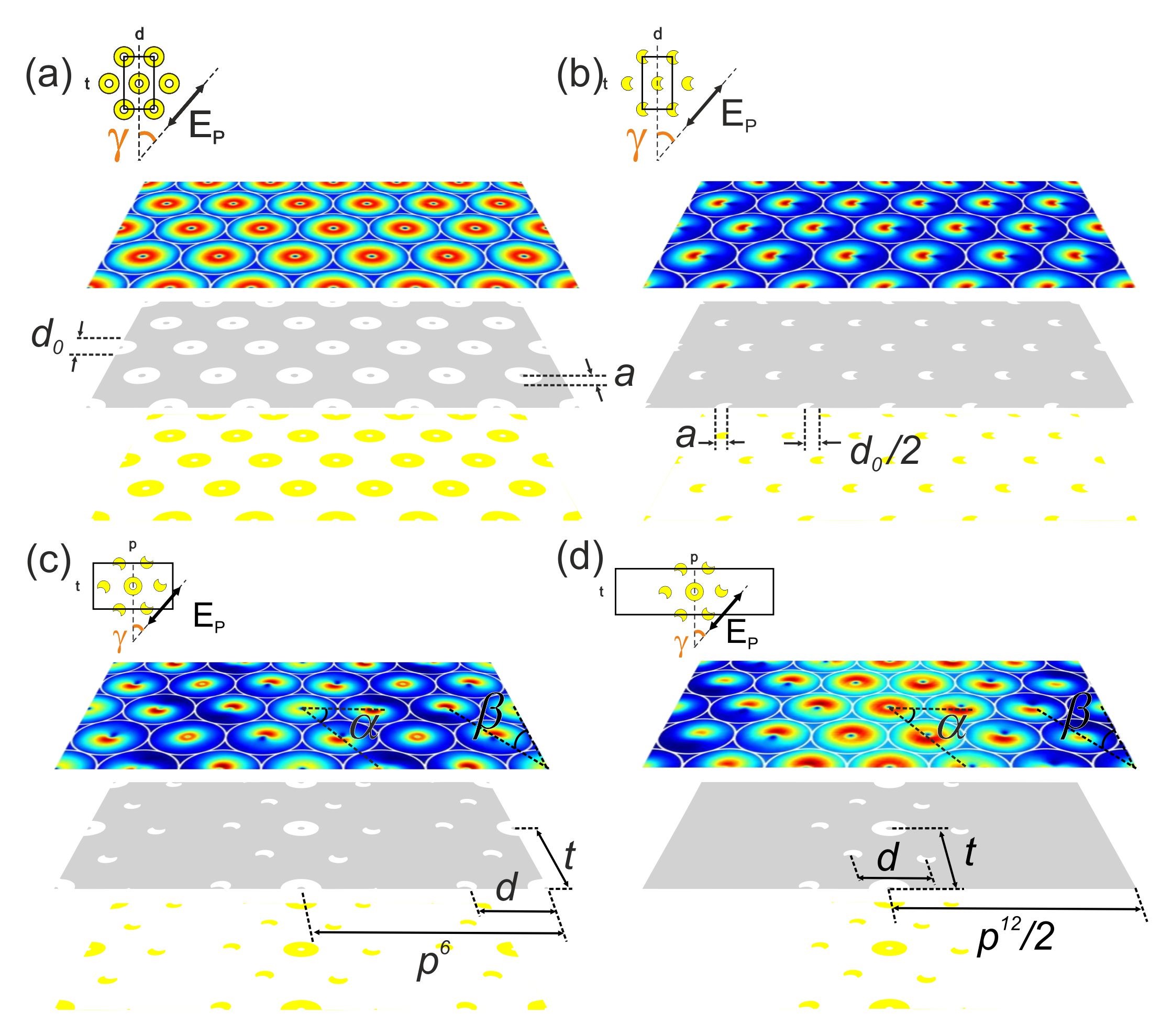}}
\caption{Normalized $\bar{E}$-field on the surface of the multilayer to be structured under Au colloid sphere monolayers illuminated by a single circularly polarized $\lambda$=400 nm beam incident (a) perpendicularly and (b) obliquely, and (c, d) by applying IICSM configuration. The insets show the unit cells of the resulted convex patterns transferred into the gold layer: hexagonal pattern of (a) nanorings and (b) nanocrescents, rectangular pattern of ring-crescent miniarrays with (c) $p$=300 nm $(n=6)$ and (d) $p'$=600 nm $(n=12)$ periodicity, by using two-beam interference pattern that is parallel to the (100) monolayer lattice direction in IICSM realization in both cases.}
\label{colloid_fig_02}
\end{figure}

Via illumination of a monolayer consisting of 100 nm Au colloid spheres by a homogeneous perpendicularly and obliquely ($\theta^6=41.8^{\circ}$) incident circularly polarized $\lambda$= 400 nm beam, hexagonal array of uniform nanorings (Fig. \ref{colloid_fig_02}a) and nanocrescents (Fig. \ref{colloid_fig_02}b) can be fabricated. Via IICSM methodology rectangular patterns composed of analogue miniarrays consisting of a central nanoring and satellite nanocrescents with $p^6$=300 nm and $p^{12}$=600 nm periodicity can be generated by two interfering circularly polarized beams incident at $\theta^6=41.8^{\circ}$ and $\theta^{12}=19.5^{\circ}$ angles, corresponding to $n$ = 6 and $n$ = 12 cases (Fig. \ref{colloid_fig_02}c, d) \cite{ref_46,ref_47,ref_48}.\\
In our present study the size parameters of the inspected ring- and crescent-shaped nano-objects, the $\varepsilon$ opening angle and the $\omega$  orientation of the nanocrescents, as well as the pattern periods are equal to those of the complementary concave patterns described in our corresponding papers, in order to ensure comparability \cite{ref_50,ref_51}.  Namely, uniform nanorings with 10 nm and 46 nm (50 nm) inner and outer diameters were inserted into the hexagonal pattern (artificial rectangular pattern of singlet nanorings and both rectangular patterns of analogue miniarrays). The nanocrescents were approximated as the intersections of two cylindrical objects with 25 nm and 20 nm diameter, with 12.5 nm center-distance. 

\subsection{Spectral and near-field effects of different patterns}
\label{Spec_nea_fie_eff}

\subsubsection{Hexagonal pattern of convex nanorings}
\label{Hex_rin}

\begin{figure}[h]
\center
	{\includegraphics[width=0.75\textwidth]{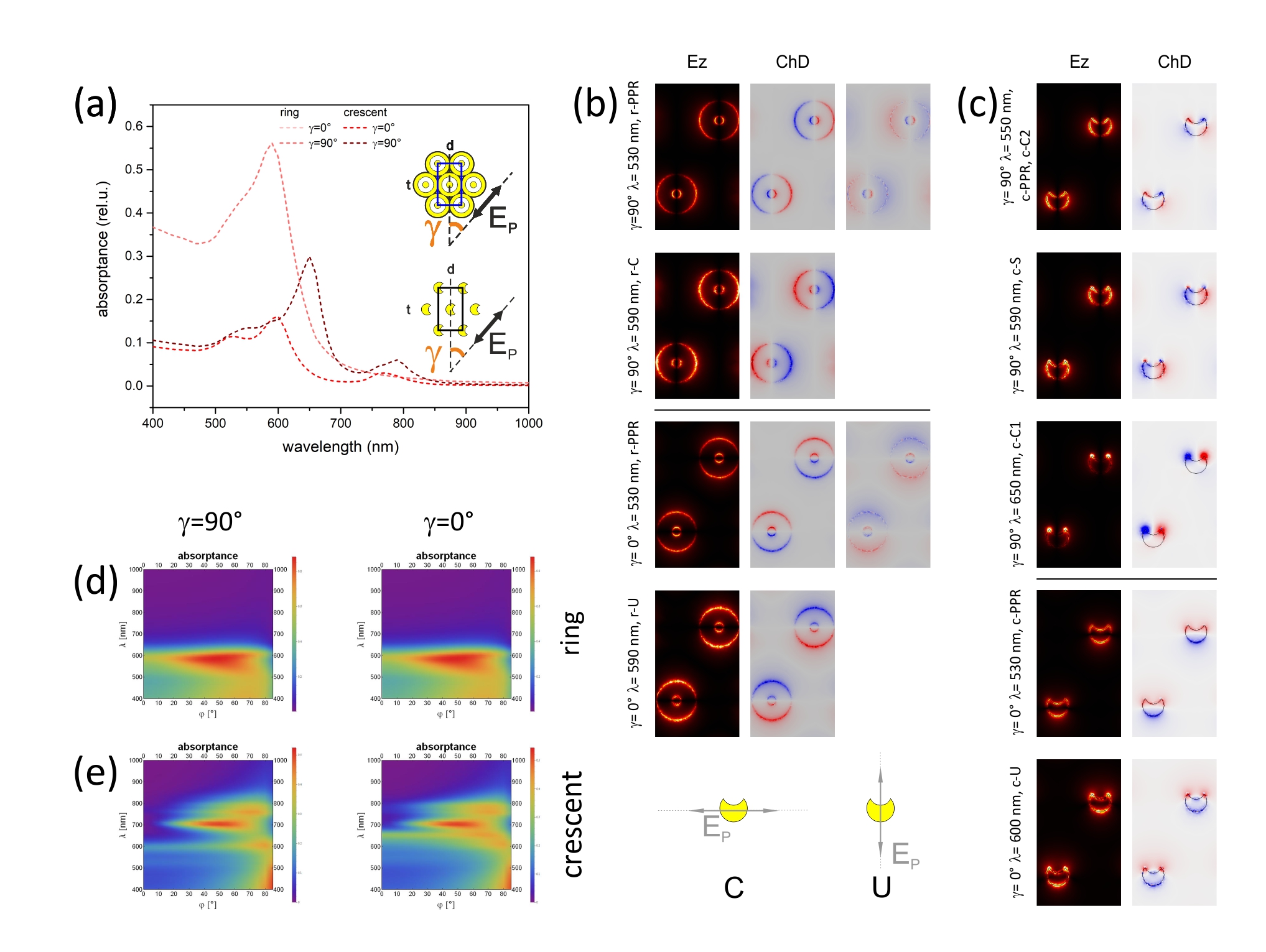}}
\caption{Hexagonal patterns composed of convex nanorings and horizontal nanocrescents: (a) absorptance spectra, $E_z$ field component and charge distribution in (top) $90^{\circ}$ and (bottom) $0^{\circ}$ azimuthal orientation of (b) nanorings and (c) nanocrescents, dispersion characteristics taken in (left) $90^{\circ}$ and (right) $0^{\circ}$ azimuthal orientation of (d) nanorings and (e) nanocrescents. Insets: schematic drawings of the unit cells and C and U configurations.}
\label{IICSM_DG_cx_01}
\end{figure}

In case of a hexagonal pattern composed of nanorings no orientation dependence is expected  based on the spherical symmetry of nanorings and on the symmetry properties of hexagonal lattices. As a result, the optical responses are azimuthal orientation independent, accordingly the absorptance spectra completely overlap in the two inspected mutually perpendicular azimuthal orientations, namely at $0^{\circ}$ and $90^{\circ}$ azimuthal angles.\\
On the absorptance of the hexagonal pattern composed of convex nanorings only a shoulder appears in the interval of particle plasmon resonance (r-PPR, 530 nm) in either of these inspected azimuthal orientations (Fig. \ref{IICSM_DG_cx_01}a). Not only parallel, but reversal dipoles also develop on the inner and outer rim in a noticeable fraction within one cycle of the time-dependent charge distribution, which correlates with the allowed different charge separations that accompany the PPR phenomenon on tiny gold nano-objects. The $E_z$ field component is relatively weaker on the outer rim of nanorings in the spectral interval corresponding to r-PPR. After the shoulder a global maximum appears at 590 nm both in $90^{\circ}$ and $0^{\circ}$ azimuthal orientations, where exclusively parallel dipoles develop on the inner and outer rim of the ring along the $\bar{E}$-field oscillation direction. These are nominated as r-C and r-U resonance in order to ensure comparability with resonances on nanocrescents. The $E_z$ field component distribution indicates more commensurate lobes both on the inner and outer rims parallel to the $\bar{E}$-field oscillation direction, both in $0^{\circ}$ and $90^{\circ}$ azimuthal orientations (Fig. \ref{IICSM_DG_cx_01}b).

\subsubsection{Hexagonal pattern of horizontal convex nanocrescents}
\label{Hex_cre}

\paragraph{Hexagonal pattern of horizontal convex nanocrescents in C orientation}
\label{Hex_cre_C}

 \ \\ 
On the absorptance of the hexagonal pattern composed of horizontal convex nanocrescents a local maximum (550 nm), a shoulder (590 nm) and a global maximum (650 nm) appears in $90^{\circ}$ azimuthal orientation, where the $\bar{E}$-field oscillation direction is perpendicular to the symmetry axis of the nanocrescents, therefore it is nominated as C orientation in accordance with the literature \cite{ref_23} (Fig. \ref{IICSM_DG_cx_01}a). The charge distribution is quadrupolar at the local maximum originating from the spectrally overlapping and interacting c-PPR and c-C2 resonance. Accordingly, there are four lobes on the $E_z$ field component distribution. At the shoulder a hexapolar charge distribution is observable, which is accompanied by a stronger charge accumulation at the tips. As a result, four weak and two bright lobes are distinguishable with intensity maxima on the nanocrescent tips on the $E_z$ field component distribution. At the global maximum dipoles arise along the $\bar{E}$-field oscillation direction, which unambiguously reveals the c-C1 resonance. Exclusively the two tips of the nanocrescents are shiny on the $E_z$ field component distribution at this maximum (Fig. \ref{IICSM_DG_cx_01}c, top). Both the charge and the $E_z$ field component distribution indicate that the shoulder originates from the interaction of the quadrupolar and dipolar modes excitable at the neighbouring extrema. Namely, the four weaker and two brighter lobes inherited from c-C2 and c-C1 resonance create two separated composite lobes with their centres at the nanocrescent tips. The small local maximum after the global maximum originates from a resonance that can be excited azimuthal orientation dependently on the tips of the nanocrescents. Both the strength and the exact location of these extrema strongly depend on the mesh density, therefore a mesh quality ensuring separation from the well-defined LSPRs has been selected. Further details are described in the Supplementary material.

\paragraph{Hexagonal pattern of horizontal convex nanocrescents in U orientation}
\label{Hex_cre_U}
 \ \\ 
In comparison, on the absorptance of the hexagonal pattern composed of horizontal convex nanocrescents a local maximum (530 nm) is followed by the global maximum (600 nm) in $0^{\circ}$ azimuthal orientation, where the $\bar{E}$-field oscillation direction is along the symmetry axis of the nanocrescents, therefore it is nominated as U orientation in accordance with the literature \cite{ref_23} (Fig. \ref{IICSM_DG_cx_01}a). Dipolar charge distribution is observable already at the local maximum corresponding to the particle plasmon resonance (c-PPR). Accordingly, there are two lobes on the $E_z$ field component distribution. At the global maximum dipoles arise along the $\bar{E}$-field oscillation direction with stronger charge accumulation on the tips, which unambiguously reveals the c-U resonance (Fig. \ref{IICSM_DG_cx_01}c, bottom). \\
The charge distribution is rearranged with respect to the c-PPR in such a way that the two tips and the long arch of the nanocrescents are the shiniest on the $E_z$ field component distribution at the global maximum. One has to emphasize that at both maxima the charge distribution is monopolar (dipolar) in 60\% (40\%) fraction of one cycle of the time-dependent charge distribution (Fig. \ref{IICSM_DG_cx_01}c, bottom).

\subsubsection{Rectangular 300 nm periodic pattern of singlet convex nanorings}
\label{Rec_rin}

\begin{figure}[h]
\center
\switchocg{imgA02 imgB02}{
  \makebox[0pt][l]{
    \begin{ocg}{Image A02}{imgA02}{on}
      \includegraphics[scale=0.6]{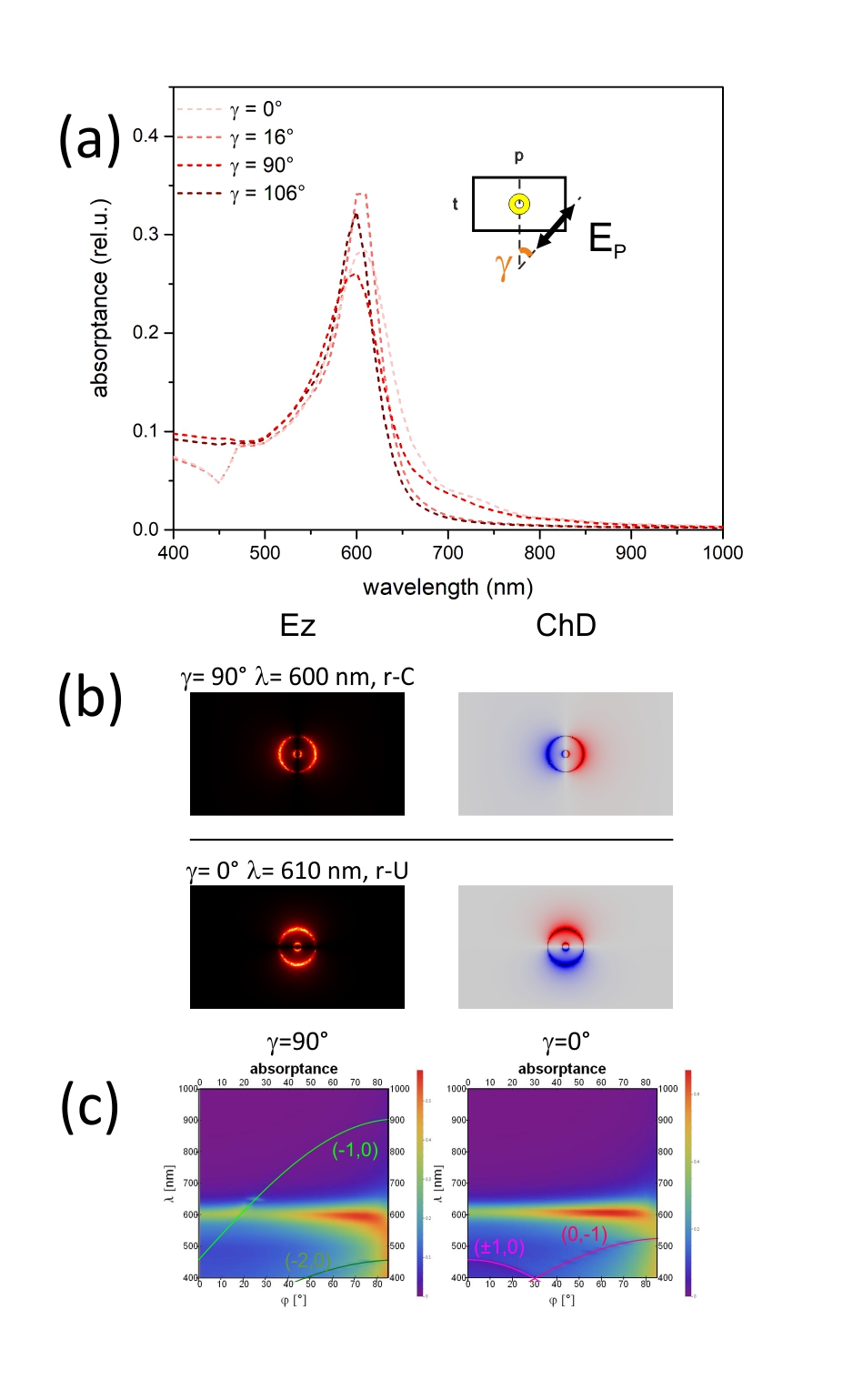}
    \end{ocg}
  }
  \begin{ocg}{Image B02}{imgB02}{off}
     \includegraphics[scale=0.6]{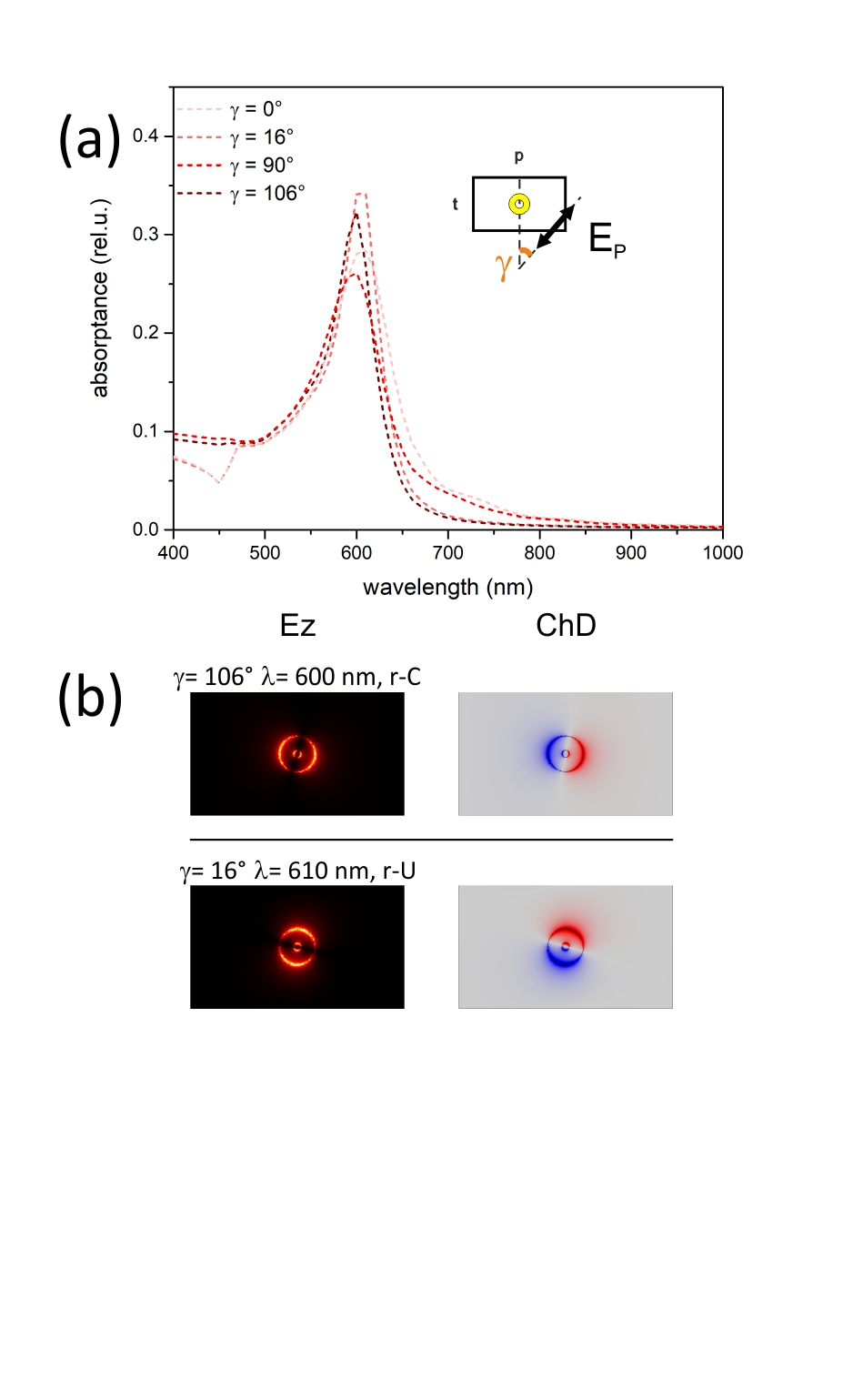}
  \end{ocg}
}
\caption{Rectangular $p$=300 nm periodic pattern composed of singlet convex nanorings: (a) absorptance spectra, (b) $E_z$ field component and charge distribution in (top) $90^{\circ}$/$106^{\circ}$ and (bottom) $0^{\circ}$/$16^{\circ}$ azimuthal orientation, (c) dispersion characteristics taken in (left) $90^{\circ}$ and (right) $0^{\circ}$ azimuthal orientation. Inset: schematic drawing of the unit cell. (Please note that by clicking on the figure you can switch between $0^{\circ}$ and $16^{\circ}$ as well as $90^{\circ}$ and $106^{\circ}$.)}
\label{IICSM_DG_cx_02}
\end{figure}

\paragraph{Rectangular 300 nm periodic pattern of singlet convex nanorings in C orientation of nanocrescents}
\label{Rec_rin_C}
 \ \\ 
When singlet convex nanorings of similar geometry as those inspected in the hexagonal pattern are arranged into a 300 nm rectangular pattern, on their absorptance no extremum / a tiny shoulder (- / 530 nm) and a global maximum (600 nm / 600 nm) appears in $90^{\circ}$/$106^{\circ}$ azimuthal orientation, which is the C orientation of horizontal singlet / slightly rotated quadrumer nanocrescents (Fig. \ref{IICSM_DG_cx_02}a). \\
The charge distribution is dipolar at the tiny shoulder corresponding to the particle plasmon resonance (r-PPR, not shown). At the global maximum parallel dipoles develop on the inner and outer rim and the $E_z$ field component lobes are aligned along the $\bar{E}$-field oscillation direction. This peak originates from the r-C resonance on the convex nanoring, which is expected to be insensitive to the $\bar{E}$-field oscillation direction based on the spherical symmetry of the convex singlet nanoring (Fig. \ref{IICSM_DG_cx_02}b, top).

%our intented 2 picture swap with click (one column wide)

\paragraph{Rectangular 300 nm periodic pattern of singlet convex nanorings in U orientation of nanocrescents}
\label{Rec_rin_U}
 \ \\ 
In comparison, on the absorptance of the 300 nm periodic rectangular pattern composed of singlet convex nanorings no extremum / a tiny shoulder (-/ 530 nm) and a global maximum (610 nm / 610 nm) appears in $0^{\circ}$ / $16^{\circ}$ azimuthal orientation, which is the U orientation of horizontal singlet / slightly rotated quadrumer nanocrescents (Fig. \ref{IICSM_DG_cx_02}a). The charge distribution is dipolar at the tiny shoulder corresponding to the particle plasmon resonance (r-PPR, not shown). At the global maximum parallel dipoles arise on the inner and outer rim and the $E_z$ field component enhancement maxima are aligned along the $\bar{E}$-field oscillation direction, which reveal the r-U resonance on the convex singlet nanoring (Fig. \ref{IICSM_DG_cx_02}b, bottom).

\subsubsection{Rectangular 300 nm periodic pattern of horizontal singlet convex nanocrescents}
\label{Rec_cre}

\begin{figure}[h]
\center
\switchocg{imgA03 imgB03}{
  \makebox[0pt][l]{
    \begin{ocg}{Image A03}{imgA03}{on}
      \includegraphics[scale=0.6]{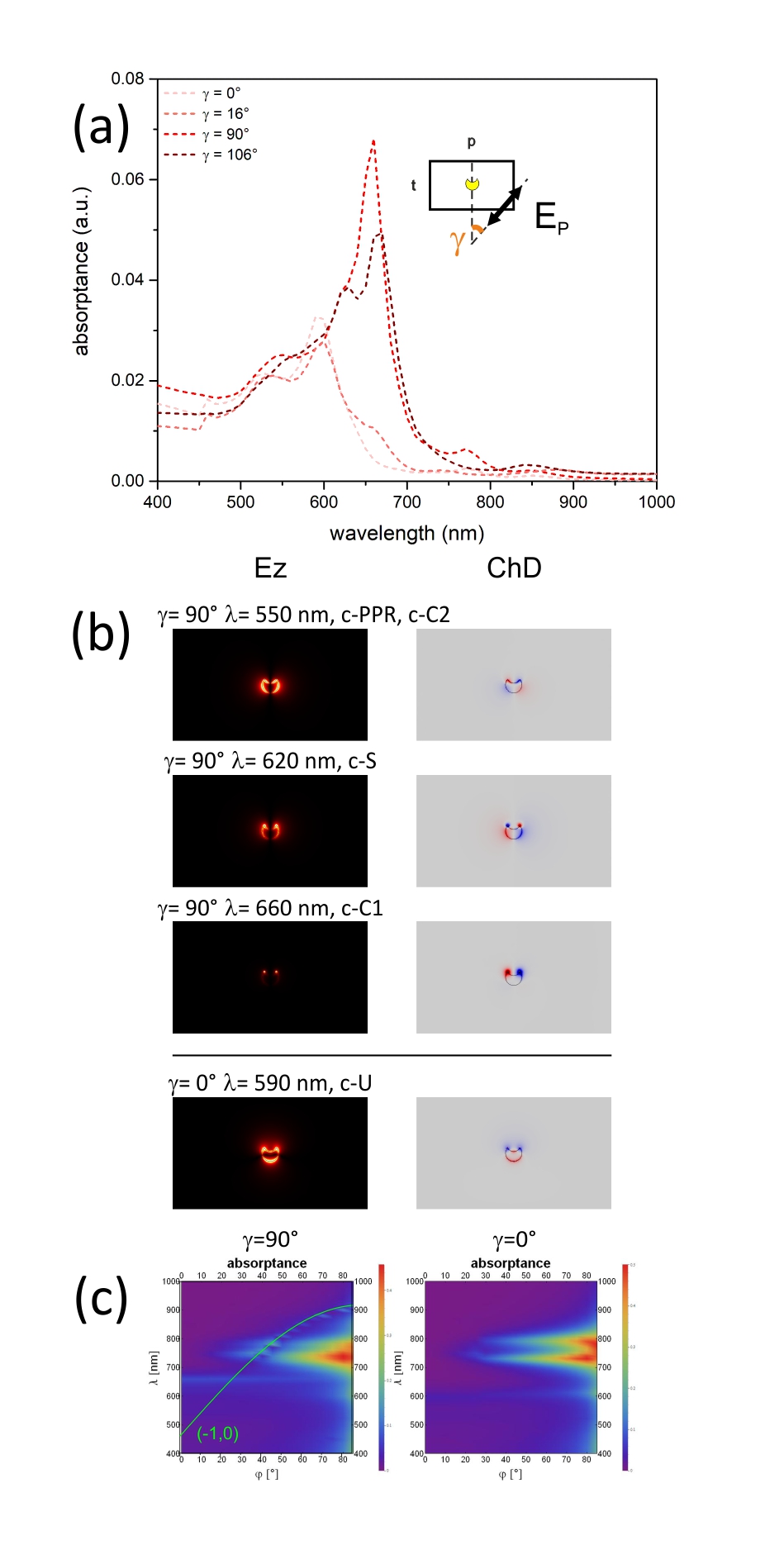}
    \end{ocg}
  }
  \begin{ocg}{Image B03}{imgB03}{off}
     \includegraphics[scale=0.6]{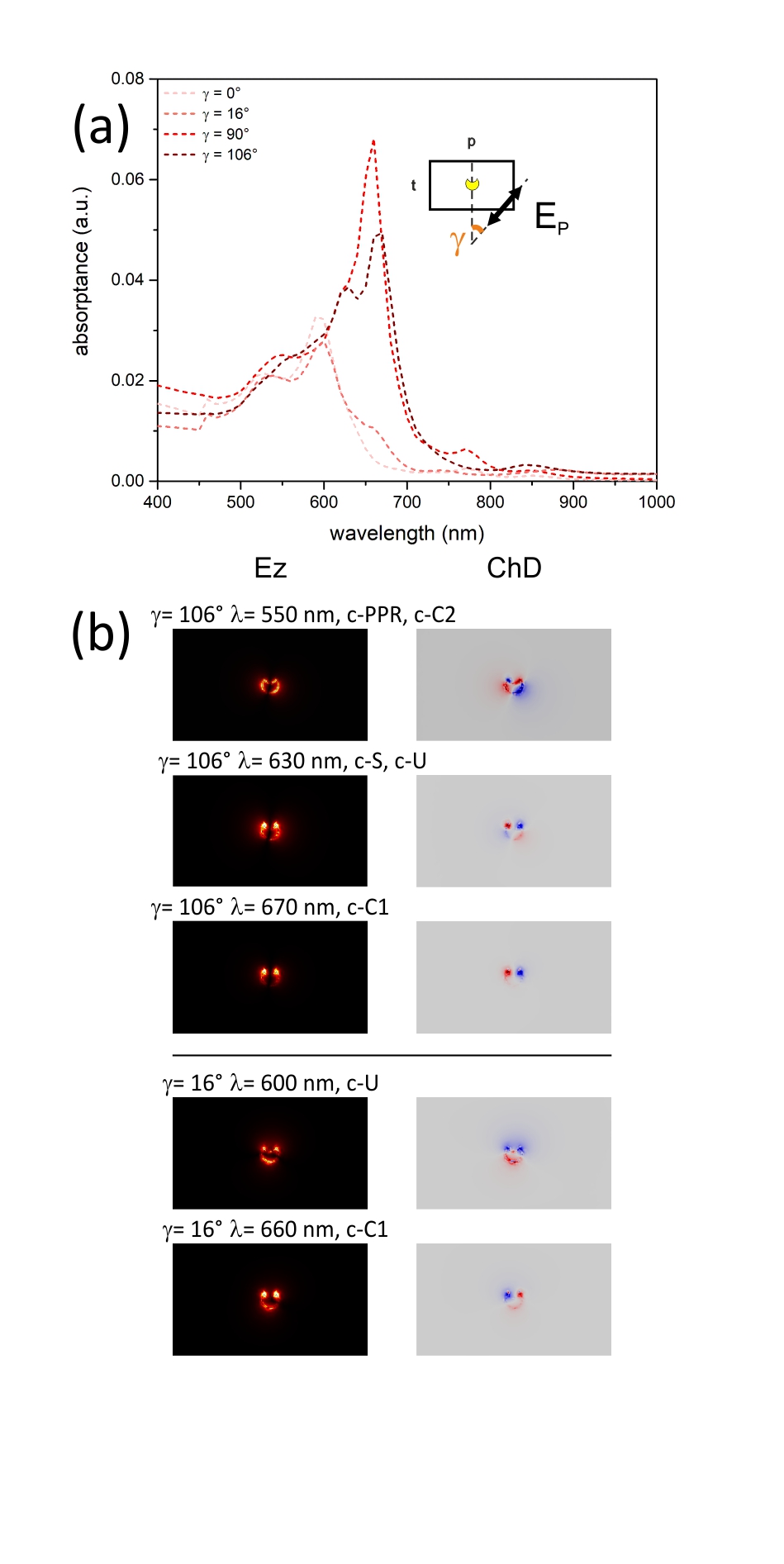}
  \end{ocg}
}
\caption{Rectangular $p$=300 nm periodic pattern composed of horizontal singlet convex nanocrescents: (a) absorptance spectra, (b) $E_z$ field component and charge distribution in (top) $90^{\circ}$/$106^{\circ}$ and (bottom) $0^{\circ}$/$16^{\circ}$ azimuthal orientation, (c) dispersion characteristics taken in (left) $90^{\circ}$ and (right) $0^{\circ}$ azimuthal orientation. Inset: schematic drawing of the unit cell. (Please note that by clicking on the figure you can switch between $0^{\circ}$ and $16^{\circ}$ as well as $90^{\circ}$ and $106^{\circ}$.)}
\label{IICSM_DG_cx_03}
\end{figure}

\paragraph{Rectangular 300 nm periodic pattern of horizontal singlet convex nanocrescents in C orientation}
\label{Rec_cre_C}
 \ \\ 
When horizontal singlet convex nanocrescents of the same shape as that inspected in hexagonal pattern are arranged into a 300 nm rectangular pattern, on their absorptance a local maximum / modulation (550 nm/ 550 nm) is followed by a shoulder / local maximum (620 nm / 630 nm) and a global maximum (660 nm / 670 nm) appears in C orientation ($90^{\circ}$) / close to it ($106^{\circ}$) (Fig. \ref{IICSM_DG_cx_03}a). \\
The charge distribution is quadrupolar at the local maximum appearing slightly above the particle plasmon resonance, which is the c-C2 resonance on the horizontal singlet convex nanocrescent in $90^{\circ}$ azimuthal orientation. Accordingly, there are four lobes on the $E_z$ field component distribution. At the shoulder / local maximum the charge distribution is still hexapolar / quadrupolar in $90^{\circ}$ / $106^{\circ}$ azimuthal orientation however the charge accumulation is stronger at the tips. Accordingly, two bright and four/two weak $E_z$ field component lobes are observable, with intensity maxima at the tips. At the global maximum dipoles arise on the nanocrescent tips along the $\bar{E}$-field oscillation direction, which reveals the c-C1 resonance on the horizontal singlet convex nanocrescent in $90^{\circ}$ azimuthal orientation. The $E_z$ field component indicates two separated lobes on the nanocrescent tips (Fig. \ref{IICSM_DG_cx_03}b, top). \\
The shoulder in C orientation originates from the interference of the c-C2 and c-C1 modes at $90^{\circ}$ azimuthal angle. Due to the $\bar{E}$-field component along the symmetry axis close to C orientation the c-U resonance is cross-coupled, as a result a maximum rather than a shoulder appears at $106^{\circ}$ azimuthal orientation in the spectral interval, which overlaps with the c-U band (Fig. \ref{IICSM_DG_cx_03}a and \ref{IICSM_DG_cx_03}b, top).  

\paragraph{Rectangular 300 nm periodic pattern of horizontal singlet convex nanocrescents in U orientation}
\label{Rec_cre_U}
 \ \\ 
In comparison, on the absorptance of the rectangular pattern composed of horizontal singlet convex nanocrescents a local maximum appearing at the particle plasmon resonance (530 nm / 540 nm) is followed by a global maximum (590 nm / 600 nm) in U orientation ($0^{\circ}$) / close to it ($16^{\circ}$), in addition to this a shoulder appears at 660 nm in $16^{\circ}$ azimuthal orientation (Fig. \ref{IICSM_DG_cx_03}a). The charge distribution is dipolar already at the local maximum appearing at the particle plasmon resonance (c-PPR, not shown). At the global maximum dipoles arise on the nanocrescent tips along the $\bar{E}$-field oscillation direction, which reveals the c-U resonance on the convex singlet nanocrescent in $0^{\circ}$ azimuthal orientation. The $E_z$ field component indicates two lobes, one localized on the tips and the other on the larger arch of nanocrescents. Caused by the rearrangement of the charge distribution with respect to c-PPR, the $E_z$ field component is less intense on the smaller arc of the nanocrescent, similarly to the hexagonal pattern. At the shoulder that follows the global maximum in $16^{\circ}$ azimuthal orientation asymmetrical  dipolar charge distribution develops, moreover the charge becomes different on the two tips. The shoulder is resulted from the appearance of the $\bar{E}$-field component in $16^{\circ}$ azimuthal orientation along the tips, which cross-couples the c-C1 resonance. This is accompanied by two separated $E_z$ field lobes, one is asymmetrically arranged on the tips, whereas the other on the larger arch of the nanocrescent (Fig. \ref{IICSM_DG_cx_03}b, bottom).  

\subsubsection{Rectangular 300 nm periodic pattern of quadrumer convex nanocrescents}
\label{Rec_qua}

\begin{figure}[h]
\center
\switchocg{imgA04 imgB04}{
  \makebox[0pt][l]{
    \begin{ocg}{Image A04}{imgA04}{on}
      \includegraphics[scale=0.6]{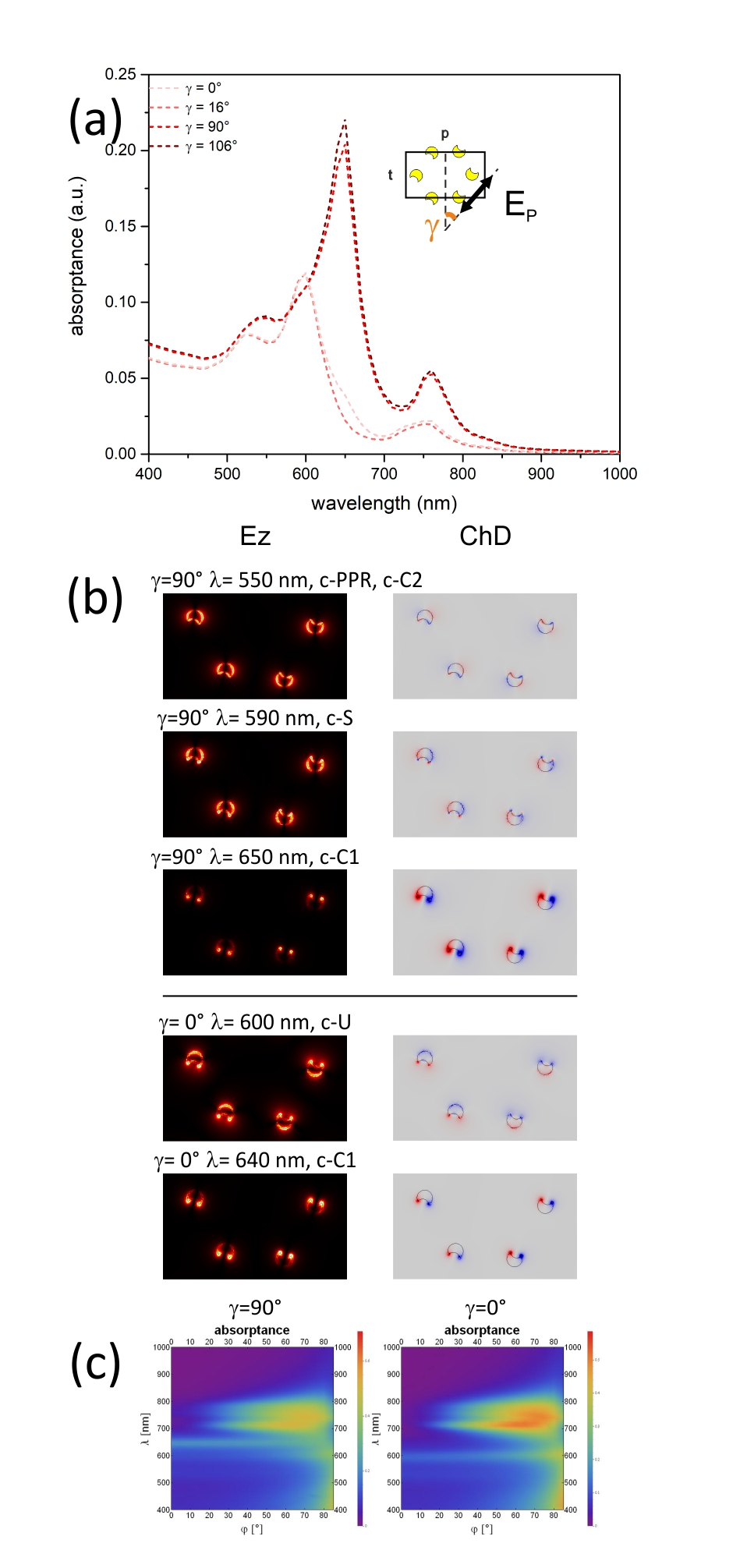}
    \end{ocg}
  }
  \begin{ocg}{Image B04}{imgB04}{off}
     \includegraphics[scale=0.6]{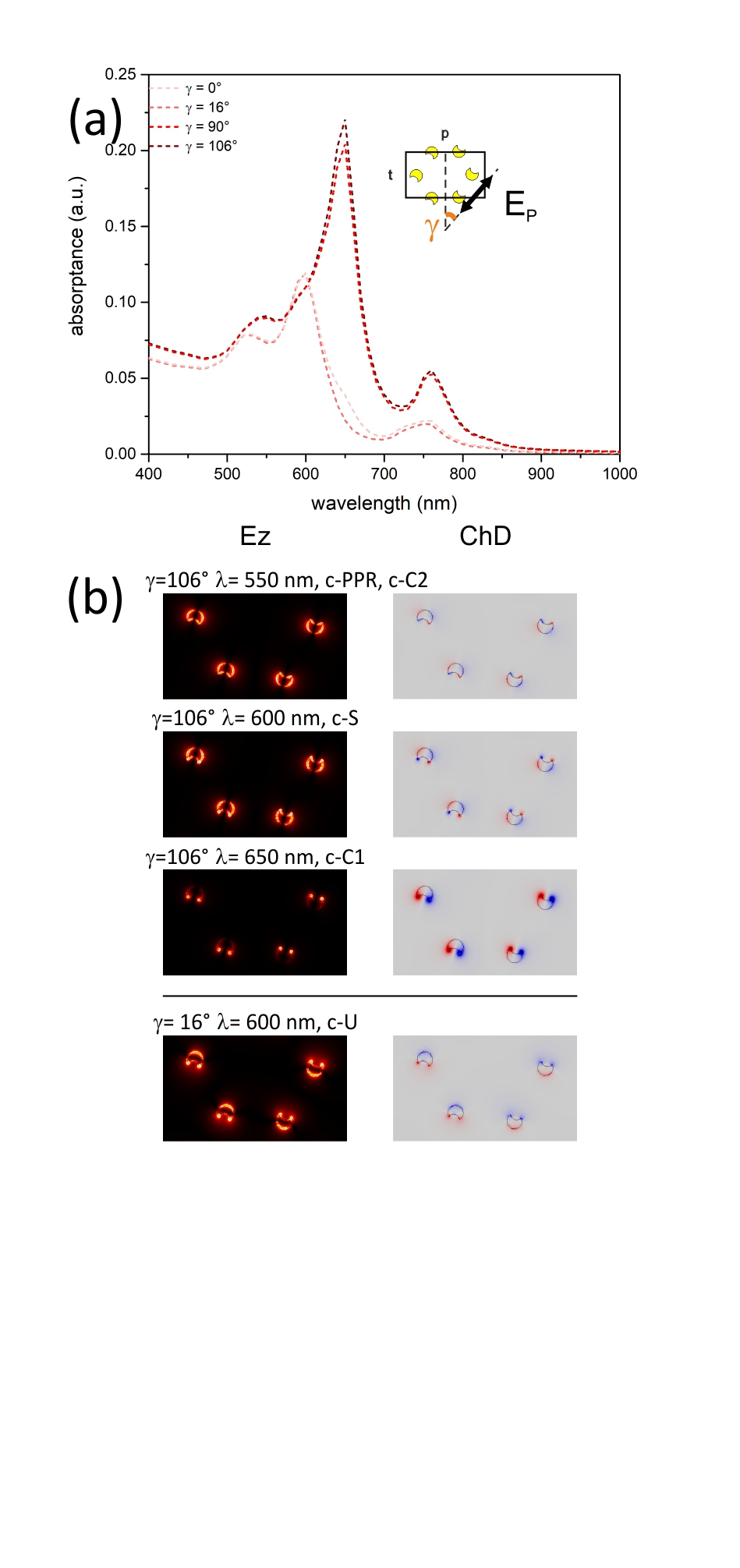}
  \end{ocg}
}
\caption{Rectangular $p$=300 nm periodic pattern composed of quadrumer convex nanocrescents: (a) absorptance spectra, (b) $E_z$ field component and charge distribution in (top) $106^{\circ}$/$90^{\circ}$ and (bottom) $16^{\circ}$/$0^{\circ}$ azimuthal orientation, (c) dispersion characteristics taken in (left) $90^{\circ}$ and (right) $0^{\circ}$ azimuthal orientation. Inset: schematic drawing of the unit cell. (Please note that by clicking on the figure you can switch between $0^{\circ}$ and $16^{\circ}$ as well as $90^{\circ}$ and $106^{\circ}$.)}
\label{IICSM_DG_cx_04}
\end{figure}

\paragraph{Rectangular 300 nm periodic pattern of quadrumer convex nanocrescents in C orientation}
\label{Rec_qua_C}
 \ \\ 
When quadrumers composed four slightly rotated convex nanocrescents are arranged into 300 nm rectangular pattern, on their absorptance a local maximum (550 nm / 550 nm) is followed by a shoulder (590 nm / 600 nm) and a global maximum (650 nm / 650 nm) appears close to ($90^{\circ}$) / in C orientation ($106^{\circ}$) (Fig. \ref{IICSM_DG_cx_04}a). Compared to rectangular pattern of singlet nanocrescents all extrema except the first are backward shifted caused by their interaction. The charge distribution is quadrupolar at the local maximum appearing slightly above the particle plasmon resonance, which reveals the c-C2 resonance on the nanocrescents quadrumer in $106^{\circ}$ azimuthal orientation. Accordingly, there are four lobes on the $E_z$ field component distribution. At the shoulder the charge distribution exhibits a hexapolar modulation / it is completely hexapolar in $90^{\circ}$/ $106^{\circ}$ azimuthal orientation. Despite the $\bar{E}$-field component appearance along the symmetry axis of slightly rotated nanocrescents close to C orientation, the c-U resonance is not efficiently cross-coupled, only a preference to quadrupolar charge separation is noticeable. The charge accumulation is stronger at the nanocrescent tips. Accordingly, two bright and four weak lobes appear on the $E_z$ field component distribution, with intensity maxima at the nanocrescent tips. At the global maximum dipoles arise on the nanocrescent tips along the $\bar{E}$-field oscillation direction, which reveals the c-C1 resonance on the convex nanocrescents quadrumer in $106^{\circ}$ azimuthal orientation. The $E_z$ field component indicates two separated lobes on the nanocrescent tips. Both the charge and the $E_z$ field distributions prove that the shoulder originates from the interference of the c-C2 and c-C1 modes (Fig. \ref{IICSM_DG_cx_04}b, top).  

\paragraph{Rectangular 300 nm periodic pattern of quadrumer convex nanocrescents in U orientation}
\label{Rec_qua_U}
 \ \\
In comparison, on the absorptance of the rectangular pattern composed of four convex nanocrescents a local maximum appearing at the particle plasmon resonance (530 nm / 530 nm) is followed by a global maximum (600 nm / 600 nm) close to ($0^{\circ}$) / in U orientation ($16^{\circ}$), in addition to this a shoulder is observable at 650 nm in $0^{\circ}$ azimuthal orientation (Fig. \ref{IICSM_DG_cx_04}a). The charge distribution is dipolar already at the local maximum appearing at the particle plasmon resonance (c-PPR, not shown). At the global maximum dipoles arise on the nanocrescent tips along the $\bar{E}$-field oscillation direction, which reveals the c-U resonance on the quadrumer of four convex nanocrescents in $16^{\circ}$ azimuthal orientation. The $E_z$ field component indicates two lobes, one asymmetrically / symmetrically arranged on the nanocrescent tips and the other on their larger arch. \\
The shoulder after the global maximum in $0^{\circ}$ azimuthal orientation originates from the cross-coupling of the c-C1 mode due to the $\bar{E}$-field component perpendicularly to the nanocrescent symmetry axis. Accordingly, the charge distribution is dipolar with different charges on the nanocrescent tips, which is accompanied by two separated lobes (Fig. \ref{IICSM_DG_cx_04}b, bottom).  

\subsubsection{Rectangular 300 nm periodic pattern of a complex convex miniarray}
\label{Min_arr_300}

\begin{figure}[h]
\center
\switchocg{imgA05 imgB05}{
  \makebox[0pt][l]{
    \begin{ocg}{Image A05}{imgA05}{on}
      \includegraphics[scale=0.6]{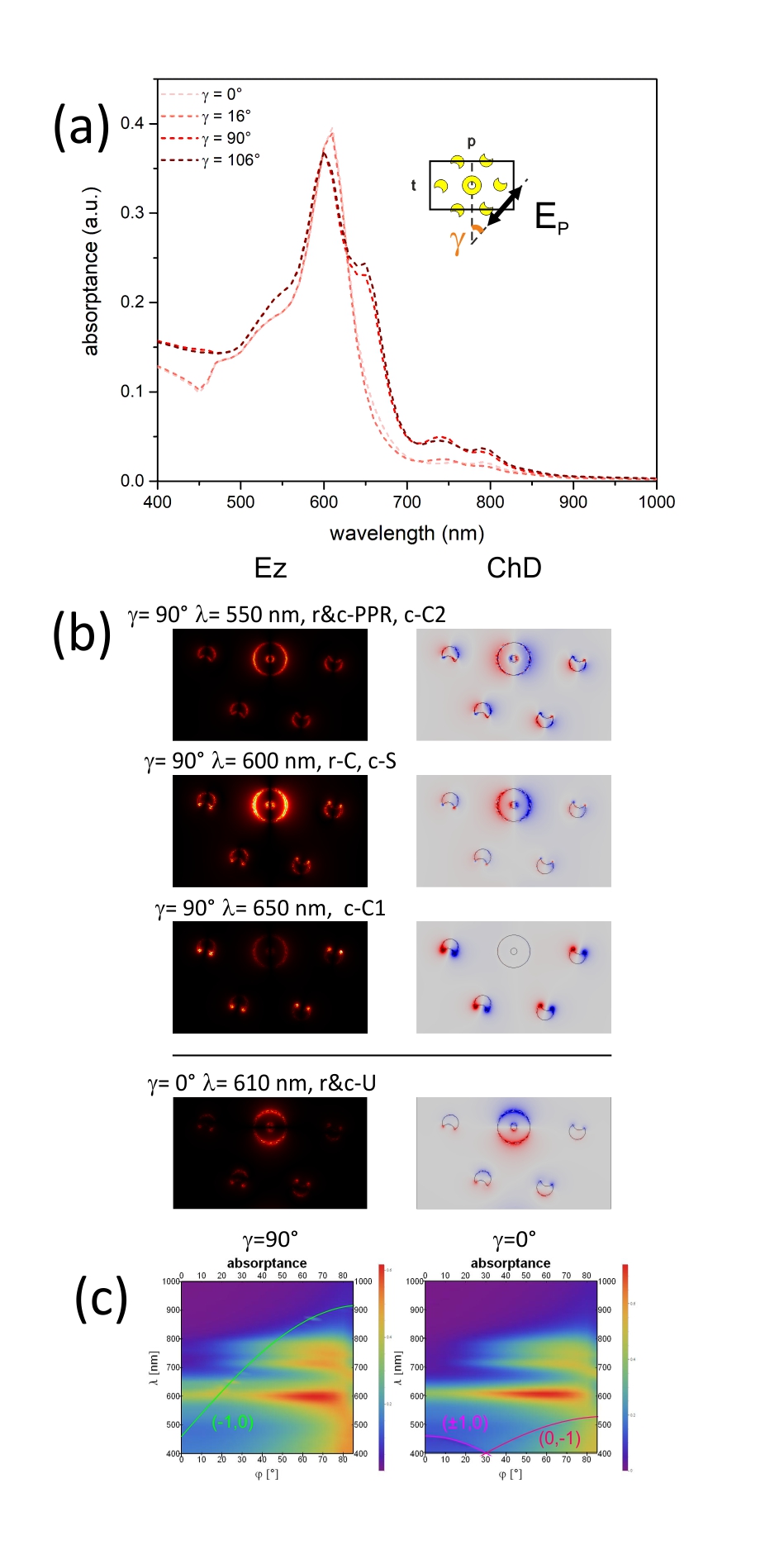}
    \end{ocg}
  }
  \begin{ocg}{Image B05}{imgB05}{off}
     \includegraphics[scale=0.6]{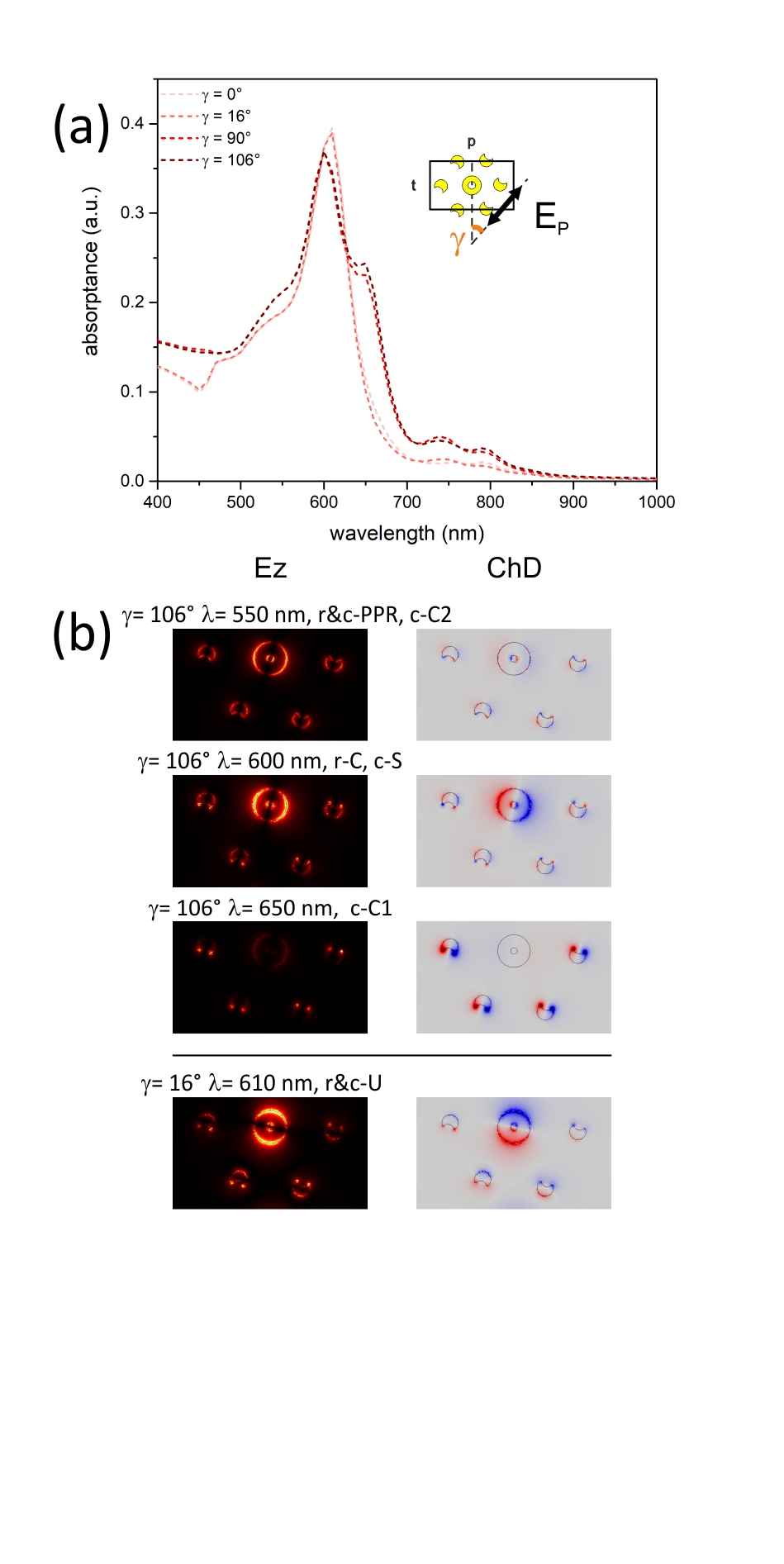}
  \end{ocg}
}
\caption{Rectangular $p$=300 nm periodic pattern composed of a complex convex miniarray: (a) absorptance spectra (b) $E_z$ field component and charge distribution in (top) $106^{\circ}$/$90^{\circ}$ and (bottom) $16^{\circ}$/$0^{\circ}$ azimuthal orientation, (c) dispersion characteristics taken in (left) $90^{\circ}$ and (right) $0^{\circ}$ azimuthal orientation. Inset: schematic drawing of the unit cell. (Please note that by clicking on the figure you can switch between $0^{\circ}$ and $16^{\circ}$ as well as $90^{\circ}$ and $106^{\circ}$.)}
\label{IICSM_DG_cx_05}
\end{figure}

\paragraph{Rectangular 300 nm periodic pattern of a complex convex miniarray in C orientation}
\label{Min_arr_300_C}
 \ \\ 
When 300 nm periodic rectangular patter is composed of a convex miniarray consisting of the central nanoring and quadrumer of nanocrescents, on their absorptance a shoulder (550 nm / 550 nm) is followed by a global (600 nm / 600 nm) and a local maximum (650 nm / 650 nm) close to ($90^{\circ}$) / in C orientation ($106^{\circ}$) (Fig. \ref{IICSM_DG_cx_05}a).  At the shoulder the charge distribution consists of reversal dipoles on the nanoring (r-PPR), whereas it is quadrupolar on the nanocrescents, which reveals the c-C2 resonance on the nanocrescents. Accordingly, there are commensurate lobes on the inner and outer rim of the nanoring, and four lobes on the nanocrescents on the $E_z$ field component distribution. \\
At the global maximum strong parallel dipoles appear on the inner and outer rim of the nanoring, whereas the charge distribution exhibits a slightly / unambiguously hexapolar modulation on the nanocrescents, which is accompanied by a stronger charge accumulation at the tips. According to the r-C ring-resonance origin of this maximum, relatively stronger $E_z$ field component lobes appear especially on the outer rim of the nanoring, whereas two bright and four weak $E_z$ field component lobes are observable on the nanocrescents, which exhibit intensity maxima on the tips that are relatively weaker compared to those on the nanoring. At the local maximum no significant charge separation occurs on the nanoring, whereas strong dipoles arise on the nanocrescent tips along the $\bar{E}$-field oscillation direction, which reveals the c-C1 resonance on the convex nanocrescent quadrumer in $106^{\circ}$ azimuthal orientation. Accordingly, the $E_z$ field component indicates week lobes on the outer rim of the ring, and two separated strong lobes on the nanocrescent tips (Fig. \ref{IICSM_DG_cx_05}b, top).\\
Comparison to rectangular pattern of a singlet nanoring and quadrumer of four nanocrescents proves that the complex miniarray response originates from the sum of the c-C2 and c-C1 modes on the quadrumer of nanocrescents, and from the r-PPR and r-C resonance of the nanoring, the former (latter) overlaps with the c-PPR (mixed mode originating from the interacting c-C2 and c-C1 resonances) on the nanocrescents (Fig. \ref{IICSM_DG_cx_05}b, top).  

\paragraph{Rectangular 300 nm periodic pattern of a complex convex miniarray in U orientation}
\label{Min_arr_300_U}
 \ \\ 
In comparison, on the absorptance of the 300 nm periodic rectangular pattern composed of a complex convex miniarray a shoulder appearing at the particle plasmon resonance (540 nm / 540 nm) is followed by a global maximum (610 nm / 610 nm) close to ($0^{\circ}$) / in U orientation ($16^{\circ}$) (Fig. \ref{IICSM_DG_cx_05}a). There are parallel / reversal dipoles on the inner and outer rim of the nanoring, whereas the charge distribution is quadrupolar / dipolar on the nanocrescents at the shoulder appearing at the PPR (r-PPR and c-PPR, not shown in Fig. \ref{IICSM_DG_cx_05}b, bottom). At the global maximum parallel dipoles arise on the inner and outer rim of the nanoring, and dipoles arise on the nanocrescents along the $\bar{E}$-field oscillation direction, which reveals the r-U and c-U resonance on the components of the miniarray in $16^{\circ}$ azimuthal orientation. \\
The $E_z$ field component indicates relatively stronger lobes on the outer rim, and two strong lobes appear on the nanocrescents, one asymmetrically / symmetrically on the tips and the other on their larger arch. The miniarray response originates from the sum of the r-PPR and r-U resonance on the nanoring and of the c-PPR and c-U modes on the quadrumer of four nanocrescents, which overlap spectrally in couples (Fig. \ref{IICSM_DG_cx_05}b, bottom).

\subsubsection{Rectangular 600 nm periodic pattern of a complex convex miniarray}
\label{Min_arr_600}

\begin{figure}[h]
\center
\switchocg{imgA06 imgB06}{
  \makebox[0pt][l]{
    \begin{ocg}{Image A06}{imgA06}{on}
      \includegraphics[scale=0.6]{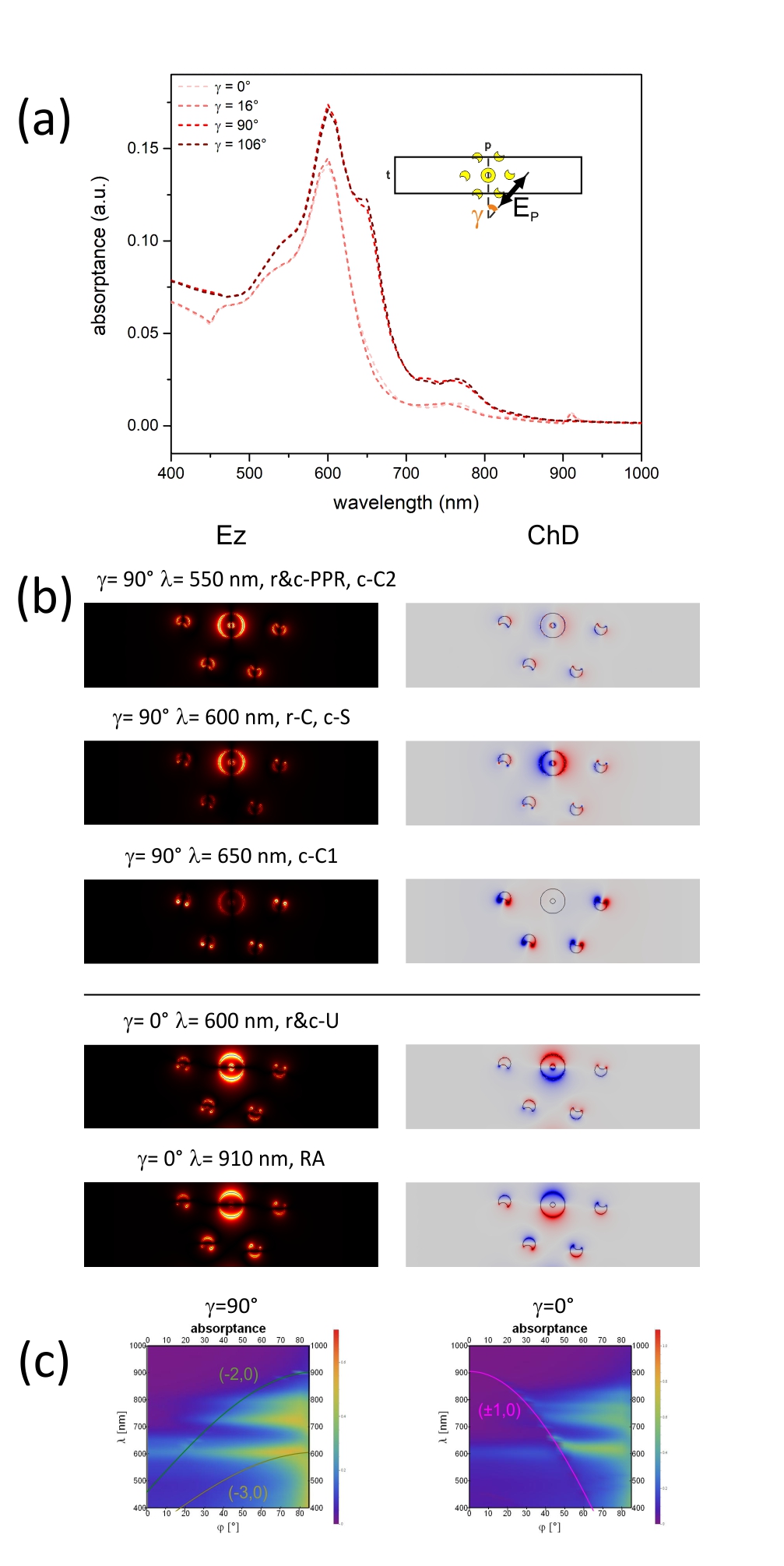}
    \end{ocg}
  }
  \begin{ocg}{Image B06}{imgB06}{off}
     \includegraphics[scale=0.6]{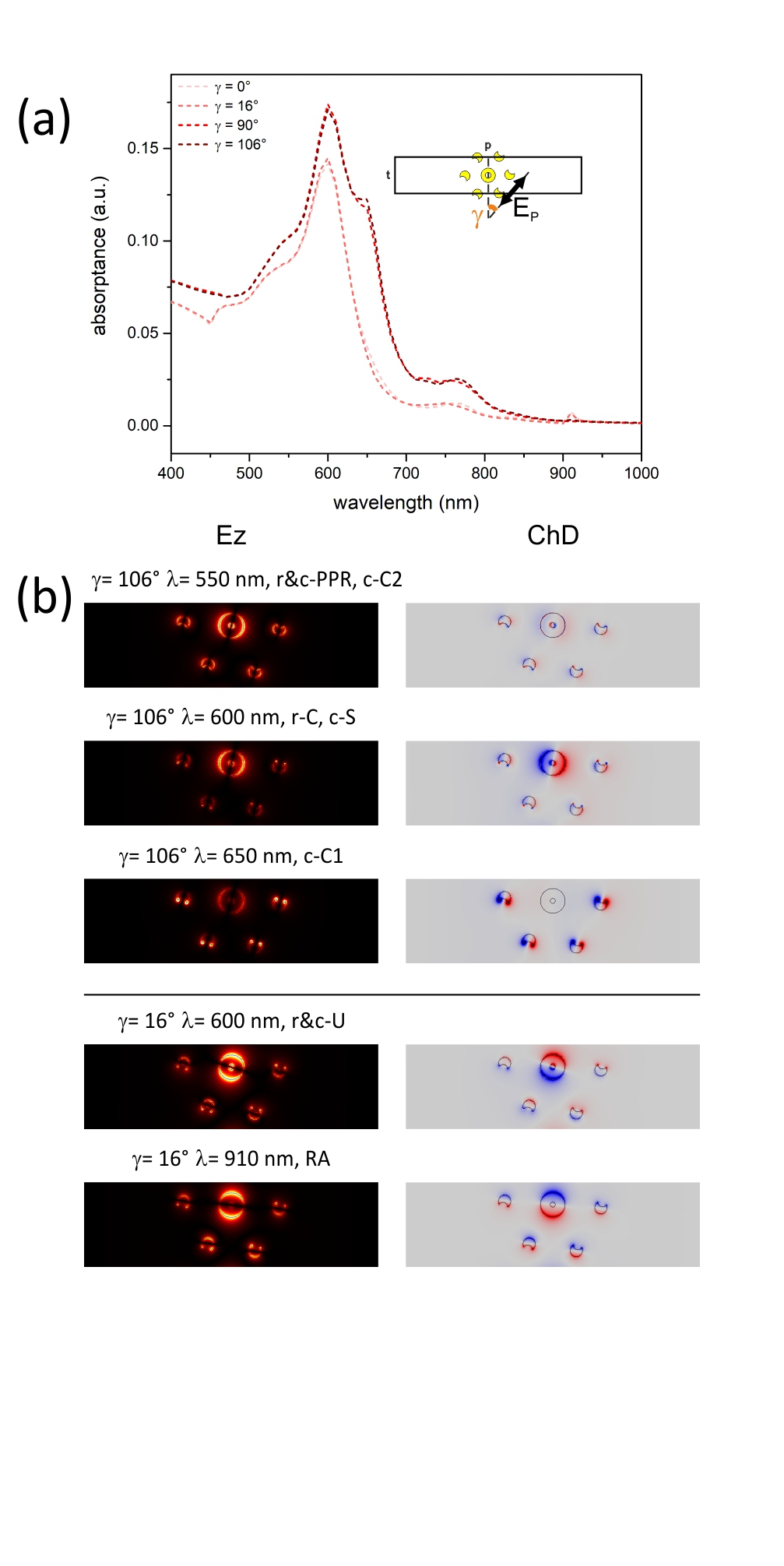}
  \end{ocg}
}
\caption{Rectangular $p'$=600 nm periodic pattern composed of a complex convex miniarray: (a) absorptance spectra, (b) $E_z$ field component and charge distribution in (left) $106^{\circ}$/$90^{\circ}$ and (right) $16^{\circ}$/$0^{\circ}$ azimuthal orientation, (c) dispersion characteristics taken in (b) $90^{\circ}$ and (c) $0^{\circ}$ azimuthal orientation. Inset: schematic drawing of the unit cell. (Please note that by clicking on the figure you can switch between $0^{\circ}$ and $16^{\circ}$ as well as $90^{\circ}$ and $106^{\circ}$.)}
\label{IICSM_DG_cx_06}
\end{figure}

\paragraph{Rectangular 600 nm periodic pattern of a complex convex miniarray in C orientation}
\label{Min_arr_600_C}
 \ \\ 
When two-times larger 600 nm periodic rectangular patter is composed of a convex miniarray consisting of the central nanoring and quadrumer of nanocrescents, on their absorptance a shoulder (550 nm / 550 nm) is followed by a global maximum (600 nm / 600 nm) and by a shoulder / local maximum (650 nm / 650 nm) close to ($90^{\circ}$) / in C orientation ($106^{\circ}$) (Fig. \ref{IICSM_DG_cx_06}a). The charge distribution consists of reversal dipoles on the nanoring (r-PPR), whereas it is quadrupolar on the nanocrescents at the shoulder, which reveals the c-C2 resonance on the miniarray in $106^{\circ}$ azimuthal orientation. Accordingly, there are commensurate lobes on the inner and outer rim of the nanoring along the $\bar{E}$-field oscillation direction, and four lobes on the nanocrescents on the $E_z$ field component distribution. At the global maximum strong parallel dipoles appear on the inner and outer rim of the nanoring along the $\bar{E}$-field oscillation direction, which reveal the r-C resonance. In contrast, in the spectral interval of former shoulders the charge distribution exhibits a slight / unambiguous mixed mode related hexapolar modulation on the nanocrescents, which is accumulated at the tips. According to the ring-origin of this maximum relatively stronger $E_z$ field component lobes appear especially on the outer rim of the nanoring along the $\bar{E}$-field oscillation direction, whereas two bright and four weak $E_z$ field component lobes are observable on the nanocrescents, which exhibit intensity maxima on the tips that are relatively weaker than those on the nanoring. At the shoulder / local maximum negligible charge separation occurs on the nanoring, whereas strong dipoles arise on the nanocrescent tips along the $\bar{E}$-field oscillation direction, which reveals the c-C1 resonance on the convex nanocrescents quadrumer in the miniarray in $106^{\circ}$ azimuthal orientation. Accordingly, the $E_z$ field component indicates significantly weaker lobes on the outer rim of the ring, and two very strong lobes on the nanocrescent tips. Similarly to 300 nm rectangular pattern, on the 600 nm periodic pattern of the analogue miniarray the optical response originates again from the sum of the c-C2 and c-C1 modes on the nanocrescents and from the r-PPR and r-C resonance of the nanoring, the former (latter) overlaps with the c-PPR (mixed mode originating from the interacting c-C2 and c-C1 modes) on the quadrumer of nanocrescents (Fig. \ref{IICSM_DG_cx_06}b, top). 

\paragraph{Rectangular 600 nm periodic pattern of complex convex miniarray in U orientation}
\label{Min_arr_600_U}
 \ \\ 
In comparison, on the absorptance of the 600 nm periodic rectangular pattern composed of a complex convex miniarray a shoulder appearing at the particle plasmon resonance (530 nm / 530 nm) is followed by a global maximum (600 nm / 600 nm), which is followed by a tiny modulation at larger wavelength (910 nm / 910 nm) close to ($0^{\circ}$) / in U orientation ($16^{\circ}$) (Fig. \ref{IICSM_DG_cx_06}a). \\
There are parallel / reversal dipoles on the inner and outer rim of the nanoring, while the charge distribution is quadrupolar / dipolar on the nanocrescents at the shoulder appearing at the particle plasmon resonance (r-PPR and c-PPR, not shown in Fig. \ref{IICSM_DG_cx_06}b, bottom). At the global maximum parallel dipoles arise on the inner and outer rim of the nanoring, and dipoles arise on the nanocrescents, all are aligned along the $\bar{E}$-field oscillation direction, which reveals the r-U and c-U resonance on the miniarray in $16^{\circ}$ azimuthal orientation. \\
The $E_z$ field component indicates relatively stronger lobes on the outer rim of the nanoring, and two strong lobes on the nanocrescents, one asymmetrically / symmetrically arranged on the tips and the other one on their larger arch. The miniarray response originates again from the sum of the r-PPR and r-U resonance of the nanoring and the c-PPR and c-U modes on the nanocrescents in quadrumer, which overlap spectrally in couples (Fig. \ref{IICSM_DG_cx_06}b, bottom).\\
At the tiny modulation related to Rayleigh anomaly, i.e. diffractive coupling outside the LSPR, the dipolar charge accumulation on the outer rim as well as on the nanocrescent tip is strengthened. Accordingly, there are strong lobes on the outer rim of the ring and strongly / intermediately asymmetrically arranged two lobes on the tips of the nanocrescents in $0^{\circ}$ / $16^{\circ}$ azimuthal orientation (Fig. \ref{IICSM_DG_cx_06}b, bottom).  

\subsection{Enhancement of dipolar emitters via rectangular patterns of different nano-objects}
\label{Enc_dip_emi}

\begin{figure}[h]
\center
	{\includegraphics[width=0.75\textwidth]{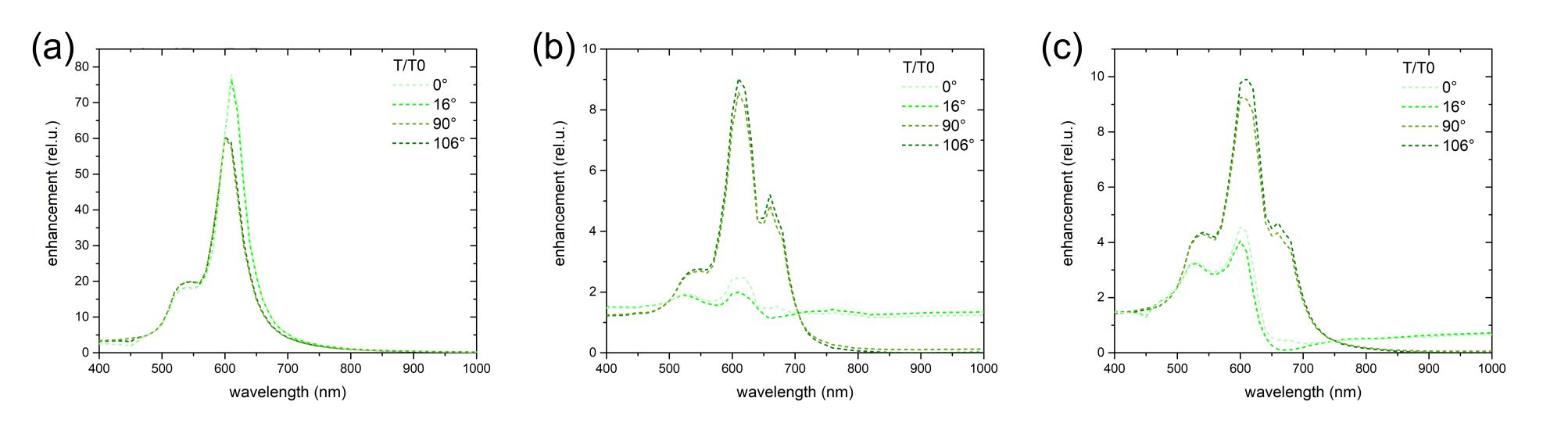}}
\caption{Radiative rate enhancement spectra of dipolar emitters achievable via 300 nm rectangular patterns consisting of convex (a) nanorings, (b) quadrumer of nanocrescents, (c) complex miniarray.}
\label{dip_encha_01}       % Give a unique label
\end{figure}

When four dipoles are deepened into the convex nanorings composing a 300 nm rectangular pattern all radiative rate enhancement spectra taken close to / in C and U orientation of quadrumer nanocrescents exhibit a r-PPR related shoulder (540 nm), and r-C and r-U mode related global maximum (600 nm and 610 nm), which are forward shifted by 10 nm with respect to their counterparts on the absorptance spectra of plane wave illuminated nanorings, except the global maximum in U orientation that is forward shifted by 20 nm (Fig. \ref{dip_encha_01}a). \\
When four dipoles are arrayed in the close proximity of four slightly rotated nanocrescents composing a quadrumer the local maxima revealing the c-C2 and c-C1 resonance close to / in C orientation appear at locations, which are coincident (550 nm) and forward shifted by 10 nm (660 nm) with respect to counterpart extrema on the absorptance spectra of plane wave illuminated quadrumers, respectively. An important difference is that a global maximum appears instead of the shoulder at a by 20 nm forward shifted location (610 nm), which originates from a unique charge hybridization. Namely, when dipoles oscillate in close proximity of the nanocrescents, reversal strongly localized and more extended dipoles are coincident on their tips (not shown). In U orientation c-PPR (520 nm) and c-U resonance (610 nm) related local and global maximum appears, which is backward and forward shifted by 10 nm with respect to counterpart extrema on the absorptance spectra of the plane wave illuminated analogue pattern, respectively (Fig. \ref{dip_encha_01}b). \\

In case of the rectangular pattern composed of a complex miniarray the peaks on the enhancement spectra are added but almost inherit the shape of the ring and quadrumer spectra, which reveals that a weak interaction occurs between the composing convex nano-objects. Namely, local maxima appear at spectral interval corresponding to the c-C2 (540 nm) and c-C1 (660 nm) resonance on the quadrumer, whereas in between them a large global maximum appears (610 nm) which originates from the mixed c-C2 and c-C1 modes overlapped with the r-C resonance on the nanoring. These extrema are shifted by -10 / 10 / 10 nm with respect to counterpart extrema on the spectra of plane wave illuminated miniarrays, respectively. In U orientation the local maximum (530 nm) originates from the r-PPR and c-PPR, whereas the global maximum is resulted from the r-U and c-U resonance on the nanorings and on the quadrumer of nanocrescents (600 nm). The former is coincident, whereas the latter is backward shifted by 10 nm with respect to counterpart extrema on plane wave illuminated miniarrays (Fig. \ref{dip_encha_01}c).

\section{Discussion and conclusion}
\label{Dis_con}

The hexagonal pattern of nanorings and nanocrescents have been inspected in $90^{\circ}$ and $0^{\circ}$ azimuthal orientations in order to uncover the characteristic LSPRs supported by the nano-objects. In case of rectangular patterns both the $90^{\circ}$ / $106^{\circ}$ and $0^{\circ}$ / $16^{\circ}$ azimuthal orientations have been studied, that are capable of promoting LSPR and grating-coupling simultaneously / LSPR separately, depending on the unit cell composition. The common difference between the charge and near-field distributions is the clockwise rotation in $106^{\circ}$ and $16^{\circ}$ azimuthal orientation with respect to those observable at $90^{\circ}$ and $0^{\circ}$ azimuthal angles.\\
On the rectangular pattern of singlet convex nanorings only the clockwise rotation is observable. On the rectangular pattern of horizontal singlet convex nanocrescents at the shoulder hexapolar charge distribution is observable in C orientation ($90^{\circ}$ azimuthal angle), whereas the charge distribution is quadrupolar close to C orientation ($106^{\circ}$ azimuthal angle). Symmetrical / asymmetrical dipoles develop in /close to U orientation ($0^{\circ}$ / $16^{\circ}$ azimuthal angle) at the global maximum.\\
On the rectangular pattern of quadrumers at the shoulder close to C orientation ($90^{\circ}$ azimuthal angle) the charge distribution is more quadrupolar, whereas the charge distribution is purely hexapolar in C orientation ($106^{\circ}$ azimuthal angle). Close to / in U orientation ($0^{\circ}$ / $16^{\circ}$ azimuthal angle) the dipolar charge distribution is asymmetrical / symmetrical on each composing nanocrescents at the global maximum. \\
The perfect C and U resonance arises in $90^{\circ}$ ($106^{\circ}$) and $0^{\circ}$ ($16^{\circ}$) azimuthal orientation on the rectangular pattern of singlets (quadrumers), which is capable of resulting in perfect alignment of the local $\bar{E}$-field perpendicularly and along the symmetry axes of the horizontal singlet (slightly rotated quadrumer composing) nanocrescents. This explains, why more well defined hexapoles arise at the shoulder in $90^{\circ}$ ($106^{\circ}$) azimuthal orientation. The well aligned dipoles development on the singlet (quadrumer composing) nanocrescents in $0^{\circ}$ ($16^{\circ}$) azimuthal orientation can be explained by their matched orientation as well. \\
The 300 nm periodic rectangular pattern composed of a convex miniarray inherits the features of the rectangular patterns composed of the singlet nanoring and quadrumer nanocrescents. Due to the perfect alignment of the $\bar{E}$-field perpendicularly to the symmetry axes of the nanocrescents in C orientation ($106^{\circ}$ azimuthal angle), the charge and field distribution is dominant on the nanoring at the global maximum and on the nanocrescent at the local maximum complementary. Due to the perfect alignment of the $\bar{E}$-field parallel to the symmetry axes of the nanocrescents the sub-sets of nanocrescents are almost indistinguishable in U orientation ($16^{\circ}$ azimuthal angle). In contrast to this, caused by the non-perfect perpendicularity of the $\bar{E}$-field oscillation direction to the symmetry axes, the modes on the nano-objects extraordinarily interact close to C orientation (in $90^{\circ}$ azimuthal angle), as a consequence a weak charge and field accumulation is observable on the nanocrescents at the global maximum and on the nanoring at the local maximum as well. Caused by the non-perfect parallelism of the $\bar{E}$-field oscillation direction to the symmetry axes close to U orientation ($0^{\circ}$ azimuthal angle), the sub-sets of nanocrescents are distinguishable. \\
The 600 nm periodic rectangular pattern composed of a convex miniarray exhibits similar features as the 300 nm periodic pattern. The differences between the perfect and non-perfect relative orientations are slightly less well defined, which can be explained by the smaller surface fraction of the nano-objects in the unit cell. Additional phenomenon is that the coincidence of the (almost) perfect alignment of the $\bar{E}$-field along the nanocrescents symmetry axes in ($0^{\circ}$) $16^{\circ}$ azimuthal orientation and the (satisfaction) approximation of the first ($\pm$1,0) order grating-coupling condition of photonic modes through $\bar{k}_p$ lattice vector results in Rayleigh anomaly outside the spectral interval of LSPRs.\\
The dispersion characteristics have been taken in $0^{\circ}$ and $90^{\circ}$ azimuthal orientations, since these make it possible to find bands associated with coupling via grating vectors of the rectangular lattice. Only coupling via $\bar{k}_p$ results in well-defined bands in the inspected wavelength interval caused by the small $t$ common side length of the inspected unit cells.\\
The common photonic band, which interacts with the inspected localized modes at large polar angles originates from (-1, 0) and (-2, 0) order grating-coupling on the 300 nm and 600 nm periodic rectangular patterns in $90^{\circ}$ azimuthal orientation, however this causes perturbation outside the spectral interval LSPRs at perpendicular incidence (Fig. \ref{IICSM_DG_cx_02}-\ref{IICSM_DG_cx_06}/c, left).\\
The dispersion characteristics of the hexagonal and rectangular pattern of convex nanorings do not possess azimuthal orientation dependence. A well-defined and tilting independent flat band is identifiable, which corresponds to the identical r-C and r-U localized plasmon resonance on the nanorings both in $90^{\circ}$ and $0^{\circ}$ azimuthal orientations (Fig. \ref{IICSM_DG_cx_01}d, Fig. \ref{IICSM_DG_cx_02}c). \\
The dispersion characteristics of the hexagonal pattern of horizontal convex nanocrescents, 300 nm periodic rectangular pattern of horizontal singlet convex nanocrescents and quadrumer of slightly rotated convex nanocrescents, as well as the 300 nm and 600 nm rectangular pattern of the complex miniarray exhibit a well-defined azimuthal orientation dependence (Fig. \ref{IICSM_DG_cx_01}e, Fig. \ref{IICSM_DG_cx_03}-\ref{IICSM_DG_cx_06}/c). \\
In $90^{\circ}$ azimuthal orientation of the hexagonal pattern as well as of the rectangular pattern of singlet and quadrumer nanocrescents a tilting independent weak flat band originating from the overlapping c-PPR and c-C2 and a strong flat band originating from the c-C1 LSPR appears, whereas the c-C2 and c-C1 modes interaction results in a ghost flat band in between them (Fig. \ref{IICSM_DG_cx_01}e, Fig. \ref{IICSM_DG_cx_03}-\ref{IICSM_DG_cx_04}/c left). This interaction related band is less well defined in case of a horizontal singlet nanocrescent in a rectangular pattern (Fig. \ref{IICSM_DG_cx_03}c left). \\
The dispersion characteristics of the 300 nm and 600 nm periodic rectangular patterns of miniarrays are more complex (Fig. \ref{IICSM_DG_cx_05}-\ref{IICSM_DG_cx_06}/c left). Similarly to the composing nano-objects, in $90^{\circ}$ azimuthal orientation of the 300 nm and 600 nm periodic rectangular patterns of the miniarray tilting independent weak flat bands originate from the overlapping r-PPR, c-PPR and c-C2 as well as from the c-C1 LSPR on the nanocrescents, whereas the c-C2 and c-C1 modes interaction results in a flat band, which overlaps with the tilting and azimuthal orientation independent strong band originating from the r-C localized plasmon resonance on the nanoring. \\
In $0^{\circ}$ azimuthal orientation of the hexagonal pattern of horizontal convex nanocrescents, rectangular pattern of horizontal singlet nanocrescents and quadrumers (as well as on the 300 nm and 600 nm periodic pattern of the complex miniarray) the weak flat band originating from the c-PPR (as well as r-PPR) is separated from the tilting independent strong band originating from the c-U (as well as r-U) LSPR (Fig. \ref{IICSM_DG_cx_01}e, Fig. \ref{IICSM_DG_cx_03}-\ref{IICSM_DG_cx_06}/c right). In addition to this, on the 600 nm periodic rectangular pattern the Rayleigh anomaly related photonic band also appears outside the spectral interval of LSPRs, due to the more / less efficient scattering of photonic modes in ($\pm$1, 0) order on the composing nano-objects along the $\bar{k}_p$ vector corresponding to their periodic pattern in $0^{\circ} / 16^{\circ}$ azimuthal orientation. \\
Our present study proves that significant enhancement of dipolar emitters is achievable in spectral intervals of plasmonic resonances on complex convex patterns that are tunable by the integrated lithography.

\begin{acknowledgements}
This work was supported by the National Research, Development and Innovation Office (NKFIH) “Optimized nanoplasmonics” (K116362) and European Union, co-financed by the European Social Fund. “Ultrafast physical processes in atoms, molecules, nanostructures and biological systems” (EFOP-3.6.2-16-2017-00005). Á. Sipos
gratefully acknowledges the support of NKFIH PD-121170.
\end{acknowledgements}



\begin{thebibliography}{}
\bibitem{ref_01}
Wang, W., Ramezani, M., Väkeväinen, A.I., Törmä, P., Rivas, J.G., Odom, T.W.: The rich photonic world of plasmonic nanoparticle arrays. Materials Today. 21, 303–314 (2018). doi:10.1016/j.mattod.2017.09.002
\bibitem{ref_02}
Guo, R., Hakala, T.K., Törmä, P.: Geometry dependence of surface lattice resonances in plasmonic nanoparticle arrays. Physical Review B. 95, 155423 (2017). doi:10.1103/PhysRevB.95.155423
\bibitem{ref_03}
Yang, A., Hryn, A.J., Bourgeois, M.R., Lee, W.-K., Hu, J., Schatz, G.C., Odom, T.W.: Programmable and reversible plasmon mode engineering. Proceedings of the National Academy of Sciences of the United States of America. 113, 14201–14206 (2016)(a). doi:10.1073/pnas.1615281113
\bibitem{ref_04}
Lévêque, G., Martin, O.J.F.: Tunable composite nanoparticle for plasmonics. Optics Letters. 31, 2750–2752 (2006). doi:10.1364/OL.31.002750
\bibitem{ref_05}
Bohren, C.F., Huffman, D.R.: Absorption and Scattering of Light by Small Particles. Wiley (1998). doi:10.1002/9783527618156
\bibitem{ref_06}
Jensen, T.R., Schatz, G.C., Van Duyne, R.P.: Nanosphere Lithography: Surface Plasmon Resonance Spectrum of a Periodic Array of Silver Nanoparticles by Ultraviolet-Visible Extinction Spectroscopy and Electrodynamic Modeling. J. Phys. Chem. B. 103, 2394–2401 (1999). doi:10.1021/jp984406y
\bibitem{ref_07}
Ruppin, R.: Spherical and cylindrical surface polaritons in solids. In: Electromagnetic Surface Modes (A. D. Boardman, ed.). pp. 345–398. John Wiley \& Sons Ltd., New York (1982)
\bibitem{ref_08}
Oldenburg, S.J., Jackson, J.B., Westcott, S.L., Halas, N.J.: Infrared extinction properties of gold nanoshells. Appl. Phys. Lett. 75, 2897–2899 (1999). doi:10.1063/1.125183
\bibitem{ref_09}
Aizpurua, J., Hanarp, P., Sutherland, D.S., Käll, M., Bryant, G.W., García de Abajo, F.J.: Optical Properties of Gold Nanorings. Physical Review Letters. 90, 057401 (2003). doi:10.1103/PhysRevLett.90.057401
\bibitem{ref_10}
Aizpurua, J., Blanco, L., Hanarp, P., Sutherland, D.S., Käll, M., Bryant, G.W., García de Abajo, F.J.: Light scattering in gold nanorings. Journal of Quantitative Spectroscopy and Radiative Transfer. 89, 11–16 (2004). doi:10.1016/j.jqsrt.2004.05.007
\bibitem{ref_11}
Fan, J.A., Wu, C., Bao, K., Bao, J., Bardhan, R., Halas, N.J., Manoharan, V.N., Nordlander, P., Shvets, G., Capasso, F.: Self-assembled plasmonic nanoparticle clusters. Science. 328, 1135–1138 (2010)(b). doi:10.1126/science.1187949
\bibitem{ref_12}
Geraci, G., Hopkins, B., Miroshnichenko, A.E., Erkihun, B., Neshev, D.N., Kivshar, Y.S., Maier, S.A., Rahmani, M.: Polarisation-independent enhanced scattering by tailoring asymmetric plasmonic systems. Nanoscale. 8, 6021–6027 (2016). doi:10.1039/c6nr00029k
\bibitem{ref_13}
Fan, J.A., Bao, K., Wu, C., Bao, J., Bardhan, R., Halas, N.J., Manoharan, V.N., Shvets, G., Nordlander, P., Capasso, F.: Fano-like interference in self-assembled plasmonic quadrumer clusters. Nano Letters. 10, 4680–4685 (2010)(a). doi:10.1021/nl1029732
\bibitem{ref_14}
Lassiter, J.B., Sobhani, H., Fan, J.A., Kundu, J., Capasso, F., Nordlander, P., Halas, N.J.: Fano resonances in plasmonic nanoclusters: Geometrical and chemical tunability. Nano Letters. 10, 3184–3189 (2010). doi:10.1021/nl102108u
\bibitem{ref_15}
Li, G., Hu, H., Wu, L.:Tailoring Fano lineshapes using plasmonic nanobars for highly sensitive sensing and directional emission. Physical Chemistry Chemical Physics. 21, 252-259 (2019). doi: 10.1039/C8CP05779F
\bibitem{ref_16}
Kretschmann, M., Maradudin, A.A.: Band structures of two-dimensional surface-plasmon polaritonic crystals. Physical Review B - Condensed Matter and Materials Physics. 66, 1–8 (2002). doi:10.1103/PhysRevB.66.245408
\bibitem{ref_17}
Haynes, C.L., McFarland, A.D., Zhao, L., Van Duyne, R.P., Schatz, G.C., Gunnarsson, L., Prikulis, J., Kasemo, B., Käll, M.: Nanoparticle Optics: The Importance of Radiative Dipole Coupling in Two-Dimensional Nanoparticle Arrays †. J. Phys. Chem. B. 107, 7337–7342 (2003). doi:10.1021/jp034234r
\bibitem{ref_18}
Murray, W.A., Astilean, S., Barnes, W.L.: Transition from localized surface plasmon resonance to extended surface plasmon-polariton as metallic nanoparticles merge to form a periodic hole array. Physical Review B - Condensed Matter and Materials Physics. 69, 165407-1-165407–7 (2004). doi:10.1103/PhysRevB.69.165407
\bibitem{ref_19}
Chu, Y., Schonbrun, E., Yang, T., Crozier, K.B.: Experimental observation of narrow surface plasmon resonances in gold nanoparticle arrays. Applied Physics Letters. 93, 181108 (2008). doi:10.1063/1.3012365
\bibitem{ref_20}
Auguié, B., Barnes, W.L.: Collective resonances in gold nanoparticle arrays. Physical Review Letters. 101, 143902 (2008). doi:10.1103/PhysRevLett.101.143902
\bibitem{ref_21}
Shahmansouri, A., Rashidian, B.: Comprehensive three-dimensional split-field finitedifference time-domain method for analysis of periodic plasmonic nanostructures: Near- and far-field formulation. Journal of the Optical Society of America B: Optical Physics. 28, 2690–2700 (2011). doi:10.1364/JOSAB.28.002690
\bibitem{ref_22}
Nishijima, Y., Rosa, L., Juodkazis, S.: Surface plasmon resonances in periodic and random patterns of gold nano-disks for broadband light harvesting. Optics Express. 20, 11466–11477 (2012). doi:10.1364/OE.20.011466
\bibitem{ref_23}
Rochholz, H., Bocchio, N., Kreiter, M.: Tuning resonances on crescent-shaped noble-metal nanoparticles. New Journal of Physics. 9, 53 (2007). doi:10.1088/1367-2630/9/3/053
\bibitem{ref_24}
Okamoto, T., Fukuta, T., Sato, S., Haraguchi, M., Fukui, M.: Visible near-infrared light scattering of single silver split-ring structure made by nanosphere lithography. Opt. Express. 19, 7068 (2011). doi:10.1364/OE.19.007068
\bibitem{ref_25}
Gwinner, M.C., Koroknay, E., Liwei, F., Patoka, P., Kandulski, W., Giersig, M., Giessen, H.: Periodic large-area metallic split-ring resonator metamaterial fabrication based on shadow nanosphere lithography. Small. 5, 400–406 (2009). doi:10.1002/smll.200800923
\bibitem{ref_26}
Klein, M.W., Enkrich, C., Wegener, M., Soukoulis, C.M., Linden, S.: Single-slit split-ring resonators at optical frequencies: Limits of size scaling. Optics Letters. 31, 1259–1261 (2006). doi:10.1364/OL.31.001259
\bibitem{ref_27}
Lahiri, B., McMeekin, S.G., Khokhar, A.Z., De La Rue, R.M., Johnson, N.P.: Magnetic response of Split Ring Resonators (SRRs) at visible frequencies. Optics Express. 18, 3210–3218 (2010). doi:10.1364/OE.18.003210
\bibitem{ref_28}
Zhou, J., Koschny, T., Kafesaki, M., Economou, E.N., Pendry, J.B., Soukoulis, C.M.: Saturation of the magnetic response of split-ring resonators at optical frequencies. Physical Review Letters. 95, 223902 (2005). doi:10.1103/PhysRevLett.95.223902
\bibitem{ref_29}
Caglayan, H., Bulu, I., Loncar, M., Ozbay, E.: Experimental observation of subwavelength localization using metamaterial-based cavities. Optics Letters. 34, 88–90 (2009). doi:10.1364/OL.34.000088
\bibitem{ref_30}
Aydin, K., Pryce, I.M., Atwater, H.A.: Symmetry breaking and strong coupling in planar optical metamaterials. Optics Express. 18, 13407–13417 (2010). doi:10.1364/OE.18.013407
\bibitem{ref_31}
Li, X., Bian, X., Milne, W.I., Chu, D.: Fano resonance engineering in mirror-symmetry-broken THz metamaterials. Applied Physics B. 122, 95 (2016). doi: 10.1007/s00340-016-6372-5
\bibitem{ref_32}
Hanarp, P., Käll, M., Sutherland, D.S.: Optical properties of short range ordered arrays of nanometer gold disks prepared by colloidal lithography. Journal of Physical Chemistry B. 107, 5768–5772 (2003)
\bibitem{ref_33}
Yu, L., Liu, G.L., Kim, J., Mejia, Y.X., Lee, L.P.: Nanophotonic crescent moon structures with sharp edge for ultrasensitive biomolecular detection by local electromagnetic field enhancement effect. Nano Letters. 5, 119–124 (2005). doi:10.1021/nl048232+
\bibitem{ref_34}
Zhou, L., Ding, F., Chen, H., Ding, W., Zhang, W., Chou, S.Y.: Enhancement of immunoassay’s fluorescence and detection sensitivity using three-dimensional plasmonic nano-antenna-dots array. Analytical Chemistry. 84, 4489–4495 (2012). doi:10.1021/ac3003215
\bibitem{ref_35}
Blanco, L.A., García De Abajo, F.J.: Spontaneous light emission in complex nanostructures. Physical Review B - Condensed Matter and Materials Physics. 69, 205414-1-205414–12 (2004). doi:10.1103/PhysRevB.69.205414
\bibitem{ref_36}
Yang, Y., Zhen, B., Hsu, C.W., Miller, O.D., Joannopoulos, J.D., Soljačić, M.: Optically Thin Metallic Films for High-Radiative-Efficiency Plasmonics. Nano Letters. 16, 4110–4117 (2016)(b). doi:10.1021/acs.nanolett.6b00853
\bibitem{ref_37}
Kosiorek, A., Kandulski, W., Chudzinski, P., Kempa, K., Giersig, M.: Shadow nanosphere lithography: Simulation and experiment. Nano Letters. 4, 1359–1363 (2004). doi:10.1021/nl049361t
\bibitem{ref_38}
Kosiorek, A., Kandulski, W., Glaczynska, H., Giersig, M.: Fabrication of nanoscale rings, dots, and rods by combining shadow nanosphere lithography and annealed polystyrene nanosphere masks. Small. 1, 439–444 (2005). doi:10.1002/smll.200400099
\bibitem{ref_39}
Nemiroski, A., Gonidec, M., Fox, J.M., Jean-Remy, P., Turnage, E., Whitesides, G.M.: Engineering shadows to fabricate optical metasurfaces. ACS Nano. 8, 11061–11070 (2014). doi:10.1021/nn504214b
\bibitem{ref_40}
Zhu, F.Q., Fan, D., Zhu, X., Zhu, J.-G., Cammarata, R.C., Chien, C.-L.: Ultrahigh-density arrays of ferromagnetic nanorings on macroscopic areas. Advanced Materials. 16, 2155–2159 (2004). doi:10.1002/adma.200400675
\bibitem{ref_41}
McLellan, J.M., Geissler, M., Xia, Y.: Edge spreading lithography and its application to the fabrication of mesoscopic gold and silver rings. Journal of the American Chemical Society. 126, 10830–10831 (2004). doi:10.1021/ja0470766
\bibitem{ref_42}
Geissler, M., McLellan, J.M., Chen, J., Xia, Y.: Side-by-side patterning of multiple alkanethiolate monolayers on gold by edge-spreading lithography. Angewandte Chemie - International Edition. 44, 3596–3600 (2005). doi:10.1002/anie.200500421
\bibitem{ref_43}
Yang, S.-M., Jang, S.G., Choi, D.-G., Kim, S., Yu, H.K.: Nanomachining by colloidal lithography. Small. 2, 458–475 (2006). doi:10.1002/smll.200500390
\bibitem{ref_44}
Jun, Y., Yu, D., George, M.C., Braun, P.V.: Holographically defined nanoparticle placement in 3D colloidal crystals. Journal of the American Chemical Society. 132, 9958–9959 (2010). doi:10.1021/ja1023628
\bibitem{ref_45}
Vogel, N., De Viguerie, L., Jonas, U., Weiss, C.K., Landfester, K.: Wafer-scale fabrication of ordered binary colloidal monolayers with adjustable stoichiometries. Advanced Functional Materials. 21, 3064–3073 (2011). doi:10.1002/adfm.201100414
\bibitem{ref_46}
Sipos, Á., Somogyi, A., Szabó, G., Csete, M.: Plasmonic Spectral Engineering via Interferometric Illumination of Colloid Sphere Monolayers. Plasmonics. 9, 1207–1219 (2014). doi:10.1007/s11468-014-9732-1
\bibitem{ref_47}
Csete, M., Sipos, Á., Szalai, A., Szabo, G.: Theoretical study on interferometric illumination of gold colloid-sphere monolayers to produce complex structures for spectral engineering. IEEE Photonics Journal. 4, 1909–1921 (2012). doi:10.1109/JPHOT.2012.2218587
\bibitem{ref_48}
Sipos, Á., Szalai, A., Csete, M.: Integrated lithography to prepare arrays of rounded nano-objects. 
Alternative Lithographic Technologies IV. 83232E (2012) doi:10.1117/12.916403
\bibitem{ref_49}
Sipos, Á., Tóth, E., Török, A., Fekete, O. A., Szabó, G., Csete, M.: Spectral engineering via complex patterns of rounded concave and convex nanoresonators achievable via integrated lithography realized by circularly polarized light. TechConnect Briefs. 373-376 (2019).
\bibitem{ref_50}
E. Tóth, Á. Sipos, O. A. Fekete, M. Cseta: Spectral engineering via complex patterns of circular nano-object miniarrays: II concave patterns tunable by integrated lithography realized by circularly polarized light
\bibitem{ref_51}
O. A. Fekete, E. Tóth, Á. Sipos, M. Csete: Comparative study on arrays of rounded convex and concave objects achievable in integrated lithography realized by circularly polarized light
\bibitem{ref_52}
Born M., Wolf E.: Principles of optics. Pergamon Press, Oxford (1986)
\bibitem{ref_53}
Palik, E.D. ed: Handbook of optical constants of solids. Acad. Press, San Diego, Calif. (1998)
\bibitem{ref_54}
Kats, M. A. , Yu, N. , Genevet, P., Gaburro, Z., Capasso, F.: Effect of radiation damping on the spectral response of plasmonic components. Optics Express. 19, 21748-21753 (2011). doi:/10.1364/OE.19.021748









%\bibitem{ref_01}
%\bibitem{ref_02}
% etc
\end{thebibliography}


% added by arXiv: 
\typeout{get arXiv to do 4 passes: Label(s) may have changed. Rerun}
\end{document}